\documentclass[lettersize,journal]{IEEEtran}
\usepackage{amsmath,amsfonts}
\usepackage{algorithmic}
\usepackage{array}
\usepackage[caption=false,font=normalsize,labelfont=sf,textfont=sf]{subfig}
\usepackage{textcomp}
\usepackage{stfloats}
\usepackage{url}
\usepackage{verbatim}
\usepackage{authblk}
\usepackage{graphicx}
\def\BibTeX{{\rm B\kern-.05em{\sc i\kern-.025em b}\kern-.08em
    T\kern-.1667em\lower.7ex\hbox{E}\kern-.125emX}}
\usepackage{balance}

\usepackage{xspace}
\usepackage{xcolor}
\usepackage{epstopdf}
\usepackage{enumitem}
\usepackage{booktabs}
\usepackage{multirow}
\usepackage{tabularx}
\usepackage{color,soul}
\usepackage[flushleft]{threeparttable}
\usepackage[font=small,labelfont=bf]{caption}
\usepackage{hyperref}

\def\eg{\emph{e.g.}\xspace}

\def\ie{\emph{i.e.}\xspace}
\def\etal{\emph{et al.}\xspace}

\newcommand{\one}{({\em i}\/)}
\newcommand{\two}{({\em ii}\/)}

\AtBeginDocument{%
  \providecommand\BibTeX{{%
    \normalfont B\kern-0.5em{\scshape i\kern-0.25em b}\kern-0.8em\TeX}}}

\newcommand{\del}[1]{\textcolor{gray}{\st{#1}}}

\newcommand{\ruoxi}[1]{\textit{\textcolor{violet}{[Ruoxi: #1]}}}
\newcommand{\lai}[1]{\textit{\textcolor{teal}{[Lai: #1]}}}

\renewcommand{\del}[1]{\textcolor{gray}{}}
\renewcommand{\ruoxi}[1]{\textit{\textcolor{violet}{}}}
\renewcommand{\lai}[1]{\textit{\textcolor{teal}{}}}

\renewcommand{\hl}[1]{#1}

\newcommand{\vulnUndreReview}[0]{3,049\xspace}
\newcommand{\nChanges}[0]{10,360\xspace}
\newcommand{\nCVEs}[0]{443\xspace}
\newcommand{\typesOfVuln}[0]{72\xspace}
\begin{document}

\title{On Security Weaknesses and Vulnerabilities \hl{in Deep Learning Systems}}


\author{Zhongzheng Lai, Huaming Chen\IEEEauthorrefmark{1}, Ruoxi Sun, Yu Zhang, Minhui Xue, Dong Yuan\IEEEauthorrefmark{1}
\thanks{Zhongzheng Lai, Huaming Chen, Yu Zhang and Dong Yuan are with the School of Electrical and Computer Engineering, The University of Sydney}
\thanks{Ruoxi Sun and Minhui Xue are with CSIRO's Data61, Australia}
\thanks{\IEEEauthorrefmark{1}Corresponding author.}
}
\maketitle

\begin{abstract}
The security guarantee of AI-enabled software systems (particularly using deep learning techniques as a functional core) is pivotal against the adversarial attacks exploiting software vulnerabilities. However, little attention has been paid to a systematic investigation of vulnerabilities in such systems. A common situation learn\hl{ed} from the open source software community is that deep learning engineers frequently integrate off-the-shelf or open-source learning frameworks into their ecosystems. In this work, we \hl{specifically look into deep learning (DL) framework and} perform the first \textit{systematic} study of vulnerabilities in DL systems through a comprehensive analysis of identified vulnerabilities from Common Vulnerabilities and Exposures (CVE) and open-source DL tools, including TensorFlow, Caffe, OpenCV, Keras, and PyTorch. We propose a two-stream data analysis framework to explore vulnerability patterns from various databases. We investigate the unique DL frameworks and libraries development ecosystems that appear to be decentrali\hl{z}ed and fragmented. By revisiting the \hl{Common Weakness Enumeration (CWE)} List, which provides the traditional software vulnerability related practices, we observed that it is more challenging to detect and fix the vulnerabilities throughout the DL systems lifecycle. Moreover, we conduct\hl{ed} a large-scale empirical study of \vulnUndreReview DL vulnerabilities to better understand the \hl{patterns of} vulnerability and \hl{the} challenges \hl{in fixing them}. We have released the full replication package at \hl{\protect\url{https://github.com/codelzz/Vulnerabilities4DLSystem}}. We anticipate that our study can advance the development of secure DL systems.
\end{abstract}


\section{Introduction}
\IEEEPARstart{T}{he} \hl{deep learning (DL)} development has greatly advanced its application in different domains, including computer vision, natural language, and signal processing. It has now emerged as one of the most promising approaches to offer cutting-edge inference capabilities by learning and mining latent relationships from complex data sources. Recent applications develop DL algorithms to solve complex scientific problems, such as building advanced recommendation systems for business demands and enhancing intelligent cyber-physical systems for industry needs. \hl{Although} it is recogni\hl{z}ed that current software development practices have yet to meet the dynamics development challenges of such systems, software engineers are facing a more challenging situation in accurately and securely developing such systems.

One tangible cause is the unique data-driven programming paradigm of AI-enabled systems, which requires complete and accurate understanding and translation of the functional and non-functional requirements from data to models and the deployment ecosystems. During the process, enormous research and development efforts have been devoted to the development of \hl{DL} frameworks and libraries. The widely adopted DL libraries have made the integration of DL models and software systems much easier\hl{,} and thus greatly facilitate the system development. \hl{Compared} to the software engineering practice that \hl{has} been extensively investigated~\cite{liu2020using,chen2021empirical,chen2020comprehensive,lou2020understanding,wu2021empirical}, \hl{an} insufficient understanding \hl{of} security weakness in DL systems, however, would result in severe consequences once exploits occur in safety-critical systems, such as aircraft flight control and autonomous driving systems~\cite{knight2002safety}.

\begin{figure}[t]
\centering
\includegraphics[width=\linewidth]{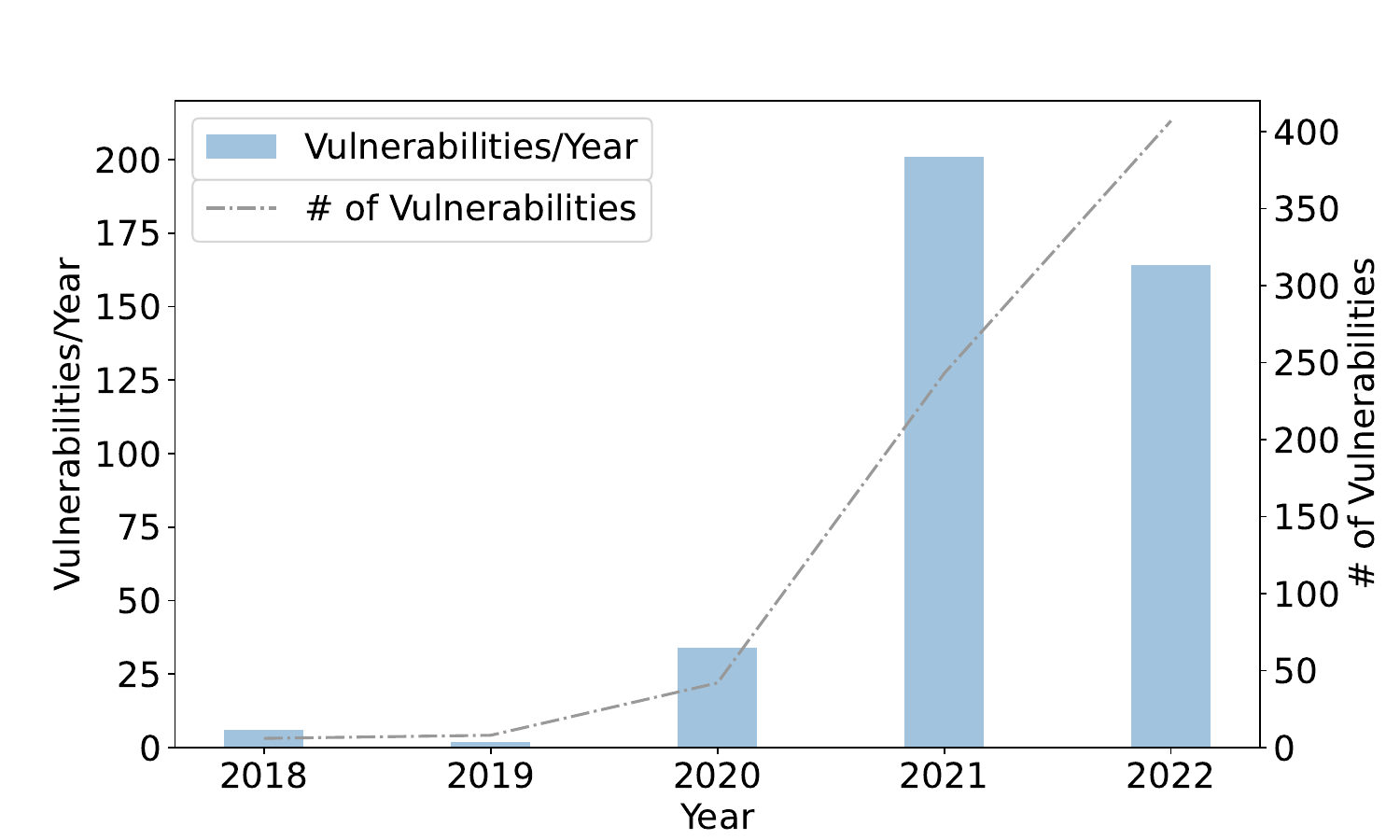}
\caption{Trend of cybersecurity vulnerabilities in TensorFlow.
}
\label{fig_trend_tf}
\vspace{-.3in}
\end{figure}

Previous work \hl{addressing} issues in DL-enabled systems has focused on multi-fold aspects, including the usage of machine learning API~\cite{wan2021machine,zhang2021unveiling}, determining and categori\hl{z}ing the bugs in \hl{DL} software~\cite{islam2019comprehensive,jia2020tfbugs,yan2021exposing,kim2021denchmark,cao2021characterizing}, and testing the \hl{DL} libraries and programs~\cite{nejadgholi2019study,pham2019cradle,wang2020deep}. However, due to limited vulnerability \hl{analysis resources}, vulnerabilities in \hl{DL} systems were much less explored in the literature than in common vulnerabilit\hl{y} analysis (such as Android system vulnerabilities~\cite{wu2019towards}). Figure~\ref{fig_trend_tf} shows the trend of disclosed cybersecurity vulnerabilities in \textit{TensorFlow}~\cite{abadi2016tensorflow}, a main-stream \hl{DL} framework, in which the \hl{number of vulnerabilities is} increasing dramatically from \textit{34} in 2020 to \textit{243} in 2021, and now \textit{407} till Jan 10th 2023. Such an exponential growth in the number of vulnerabilities is unprecedented,
which motivates us to further investigate software vulnerabilities in \hl{DL} systems.

\hl{Although} identified vulnerabilities have become overwhelming, especially in the National Vulnerability Database (NVD) with expert-verified Common Vulnerabilities and Exposures (CVE) ID, \hl{our} aim \hl{is} to investigate vulnerabilities in \hl{DL} systems. To achieve this goal, we select the five most popular \hl{DL} frameworks which are actively maintained, including \textit{TensorFlow}~\cite{abadi2016tensorflow}, \textit{Caffe}~\cite{jia2014caffe}, \textit{OpenCV}~\cite{bradski2008learning}, \textit{PyTorch}~\cite{paszke2019pytorch} and \textit{Keras}~\cite{gulli2017deep}. The reason that we chose \hl{multi-DL} open source projects rather than focusing on a particular one is to cover \hl{DL} systems as broad as possible. While only \textit{TensorFlow} maintains the \textit{Security Advisory} in the official project, this study has expanded the \hl{scope of the investigation with} widely deployed rule-based methods for vulnerability detection from relevant projects.

Specifically, we focus on vulnerabilities with \hl{open source code available} and developer discussions across project repositories. \hl{These} repositories give us full access to study \hl{vulnerabilities in terms of} description, location, and patch. We conclude the results and findings \hl{as the contributions in the following list}:

\begin{itemize}[leftmargin=*]
\item We conduct the first \textit{systematic} study of vulnerabilities in DL systems through manual analysis of \vulnUndreReview vulnerabilities from over \nChanges changes (commits/pull requests) and \nCVEs official CVEs in five DL open-source frameworks.
\item We provide an analysis of the root causes and symptoms of vulnerabilities, resulting in a classification that could benefit future research on \hl{the analysis of} DL vulnerabilities.
\item We discuss the challenges in detecting and fixing vulnerabilities in DL systems and suggest future research directions to advance the detection and \hl{patching of DL vulnerability}. 
\item We open-source the source code, datasets, and analysis results as a public replication package for software security researchers and practitioners, which can be found at~\cite{package2022investigation}.
\end{itemize}
\section{Background and Related Work}
\subsection{Background and Motivation}
\label{sec:background_motivation}
It is widely acknowledged that \hl{DL} has enhanced a growing number of areas, such as scientific research and industry application, with intelligent capability. 
Potential areas have seen efficient data-driven solutions including self-driving cars~\cite{rao2018deep}, fraud news detection~\cite{liu2020fned} and visual recognition~\cite{wang2020deep1}.
However, the development of such a DL-enabled system requires dedicated efforts \hl{from} all stakeholders, \hl{including}, but not limited \hl{to,} problem definition, requirement analysis, and model development~\cite{cysneiros2018software,bosch2021engineering,chen2020comprehensive,nikanjam2021design,giray2021software}. In most cases, it will be even difficult for developers to implement a DL program from scratch since \hl{a} decent mathematical foundation will be required~\cite{zhang2021unveiling}. Another challenge is the tedious multistage for developing DL-enabled systems~\cite{de2019understanding,john2020architecting}. Thus, enormous efforts from academia and industry have been dedicated to supporting and facilitating the development process, such as recently emerging infrastructure frameworks and services~\cite{abadi2016tensorflow,paszke2019pytorch,jia2014caffe,gulli2017deep,ye2020put,malta2019exploring}.

While the successful deployment of an intelligent system is critical, recent research also evaluates DL systems \hl{in terms of} its security and privacy~\cite{papernot2018marauder,he2020towards,evtimov2020security,chen2022security,liu2021machine,harzevili2022characterizing}. 
One important category of security threats for DL systems is introduced by infrastructure frameworks, such as TensorFlow and PyTorch~\cite{zhang2021unveiling}. Such frameworks are generally designed to offer off-the-shelf solutions and access to specific computational resources with high-level APIs. However, a first reported vulnerability related to DL systems is identified as CVE-2018-5268 in NVD. It can exploit \hl{the} Microsoft Cognitive Toolkit (CNTK) by launching remote attacks for a total loss of Confidentiality, Integrity and Availability (CIA) of the system without required authentication~\cite{cert2020,kim2020impact}. Taking TensorFlow as an example in Figure~\ref{fig_trend_tf}, the number of official vulnerabilities in NVD has greatly increased since 2021, with a total of 407. This indicates that the critical security risks in DL systems are expanding dramatically.  
It remains limbo for an in-depth and practical understanding to incorporate the knowledge from NVD, especially with the \hl{well-defined Common Weakness Enumeration (CWE) list} research concept. 

In this work, we anticipate for a large-scale systematic investigation of the vulnerabilities in DL systems with up-to-date CWE tools, particularly for DL infrastructure frameworks, to understand their root causes, symptoms\hl{,} and fixed pattern directly related to the specific DL system characteristics. Further actionable implications for all practitioners and researchers are provided with manual efforts of analysis for the unique and reliable resources, such as source code, developer analysis and discussion\hl{,} and so on. The following sections will convey the details of the methodology and the data analysis process.
To highlight the critical value of our work \hl{compared to} other most recent studies, Filus~\etal~\cite{filus2023software}initially attempts to explore 104 CVE instances related to memory operations in Tensorflow. However, only \hl{six} different CWE types are concluded as findings, which limits its general applicability for vulnerability analysis and \hl{the} specific implication for \hl{the development of} secured AI-enabled software systems. Another recent work~\cite{chen2021security} has an emphasis on research on attack and defensive approaches for DL frameworks. 

\subsection{Related Work}
\label{sec:related_work}
It is a hot topic to practically understand \hl{the process of AI-enabled system development,} particularly with advanced \hl{DL techniques}. \hl{In addition to} the work in Sec.~\ref{sec:background_motivation}, two directions are related to our works, namely \textit{empirical study of AI-enabled systems}, and \textit{testing on AI-enabled systems}.

\noindent \textbf{An Empirical Study of AI-enabled Systems.~}
For the empirical study of AI-enabled systems, we mainly discuss the existing works on identifying the latent bugs and faults for the development and deployment of such systems. It should be noted that many research challenges in relation to AI-enabled systems from the perspective of software engineering may have focused on engineering requirement analysis and \hl{the} semantic development framework. We limit the related works here for bugs and faults related empirical analysis.

The work by Thung \etal~\cite{thung2012empirical} studied three different machine learning-based systems, including Apache Mahout, Lucene, and OpenNLP. \hl{General} bugs for machine learning-based systems were surveyed with the proposed category, severity, and \hl{impact of the bug}.\hl{Zhang \protect\etal~\cite{zhang2020machine}} \hl{conducted a comprehensive survey covering 144 papers on bug testing in machine learning-based systems.} Zhang \etal~\cite{zhang2018empirical} further explored the bugs in TensorFlow-built DL applications. In total, 175 TensorFlow application bugs were collected from Stack Overflow QA pages and GitHub.\par
In ~\cite{islam2019comprehensive}, the focus was on bugs in software that \hl{use DL} libraries, while some empirical studies are subject to specific types of bugs, such as performance bugs, numerical bugs, and tensor shape faults~\cite{jia2020tfbugs,yan2021exposing,kim2021denchmark,cao2021characterizing,wu2021empirical,du2021empirical}.

\noindent \textbf{Testing on AI-enabled Systems.~}
In order to identify and evaluate the general bugs and faults in AI-enabled systems, software testing is one prominent technique. One identified challenge for testing on AI-enabled systems is defining the test oracle for the implementation of machine learning classification algorithms~\cite{xie2011testing,nejadgholi2019study}. Differential testing has also been applied to test DL libraries, such as CRADLE in ~\cite{pham2019cradle}. \hl{The test of DL} libraries by generating effective \hl{DL} models \hl{through} guided mutation is proposed as LEMON (\hl{DL} \textbf{L}ibrary t\textbf{E}sting via guided \textbf{M}utati\textbf{ON}) in~\cite{wang2020deep}. \hl{Christou \protect\etal~\cite{christou2023ivysyn}} \hl{proposed IvySyn to automatically detect DL kernel code implementations, introduce fuzzing hooks for type-aware mutation-based fuzzing, and synthesize code that propagates offending inputs for vulnerability testing.} Other works have considered \hl{the} testing for \hl{DL} from different aspects, such as prioriti\hl{z}ing inputs~\cite{wang2021prioritizing}, testing for \hl{DL} operators~\cite{zhang2021predoo,zhang2021duo}.

%
%
%

\begin{table*}[t]
\caption{Statistics of the studied deep learning frameworks in our study}\label{tab_statistics-1}
\centering
\resizebox{0.95\linewidth}{!}{
\begin{threeparttable}
\begin{tabular}{lrrrrrrr}
\toprule 
\textbf{Framework} & \textbf{GitHub Stars}  &  \textbf{Commits} &  \textbf{Pull Requests} &  \textbf{SLOC}\tnote{*} \textbf{(C/C++)} & \textbf{SLOC (Python)} &   \textbf{Closed Issues} &   \textbf{Vulnerabilities}  \\
\midrule
TensorFlow & 171.0k & 149,895 & 22,484 & 2,263,966 & 687,408 & 34,434 & 2,154\\
PyTorch    & 62.0k  & 137,429 & 62,215 & 1,207,216 & 780,704 & 20,730 & 217\\
OpenCV     & 66.1k  & 33,142  & 13,625 & 1,372,158 & 35,501  & 7,135  & 563\\
Keras      & 57.2k  & 7,886   & 5,831  & 0         & 194,240 & 11,325 & 45\\
Caffe      & 33.1k  & 12,633  & 2,240  & 61,299    & 5,379   & 3,891  & 70\\
\bottomrule
\end{tabular}
\begin{tablenotes}
\item[*] \textbf{SLOC}: source lines of code.
\end{tablenotes}
\end{threeparttable}
}
\end{table*}

\subsection{Vulnerability Management Related Database}
\label{sec:vuln_management_related_db}

\hl{Common Weakness Enumeration (CWE)}~\cite{CWE2022}\hl{, Common Vulnerabilities and Exposures (CVE)}~\cite{CVE2022}\hl{, and National Vulnerability Database (NVD)}~\cite{NVD2022}\hl{ are the three important databases that standardize the way vulnerabilities are identified, described, and addressed. CWE}~\cite{CWE2022}\hl{ is a list that contains a comprehensive classification taxonomy to identify and describe weaknesses in software and hardware systems. It focus on abstract descriptions of common weaknesses, which attempt to capture the root causes that lead to security problems. CVE}~\cite{CVE2022}\hl{ aims to identify, define, and catalog vulnerabilities. For each record, CVE provides a unique ID and a consistent description to ensure that professionals discuss and work on the same issues. NVD}~\cite{NVD2022}\hl{ is supplementary of CVE which provides enhanced information about vulnerability, including severity scores and patch availability. It collects and analyze the public available information related to each CVE and provides richer information to help professionals better understand and evaluate the issues.}
\section{Methodology}
\label{sec:methodology}
In this section, we outline the open-source data collection and analysis methods used in this work, covering the five most popular \hl{DL} frameworks (TensorFlow, Caffe, OpenCV, PyTorch and Keras). We curate, process, and analyze the data with cross-reference checking from multiple sources,
including GitHub repositories and NVD. The ultimate goal is to ensure that we have obtained a high-quality dataset to characterize and investigate vulnerabilities in the context of DL systems, along with the latest \hl{definitions of vulnerabilities} in the CWE List version 4.9~\cite{CWE2022}.

\subsection{Vulnerability Dataset Establishment}\label{sec_data_collection}
\begin{figure}[t]
\centering
\includegraphics[width=\columnwidth]{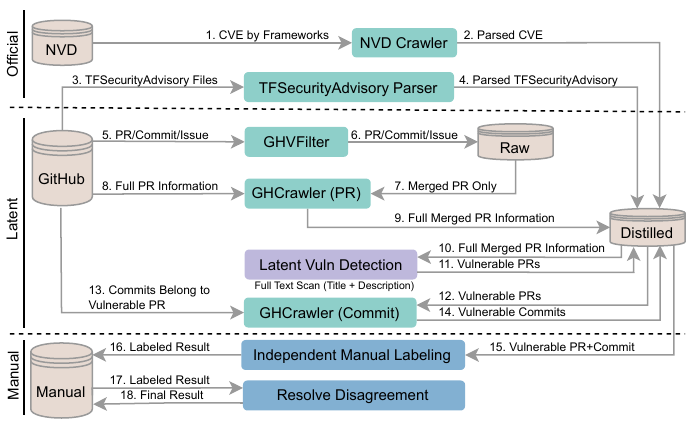}
\caption{An overview of proposed two-stream data analysis framework.}
\label{fig:data_collection}
\end{figure}

In this section, we introduce our method for \hl{collecting vulnerability datasets}, aiming to comprehensively identify all vulnerabilities in line with all possible software security related issues. Figure~\ref{fig:data_collection} shows a two-stream method obtains high-fidelity vulnerabilities in a multi-round process with a well-designed reference checking support among public CVEs, GitHub pull requests, issue reports\hl{,} and commit logs.

\noindent \textbf{Data collection.~}
In particular, one stream focuses on NVD, while another stream targets GitHub. 
In order to automatically identify vulnerabilities from GitHub repositories, we design a comprehensive set of regular expression rules deri\hl{v}ed from the existing works ~\cite{bosu2014identifying, zhou2017automated}. 
To ensure the quality and effectiveness of the search keywords, we followed a four-step empirical approach as described below:

\begin{itemize}
\item Initial keyword set building. We combine the vulnerability search keywords from~\cite{bosu2014identifying} with the regular expression from ~\cite{zhou2017automated}, establishing an initial set of keywords.
\item Pilot validation. We use the TensorFlow official ``Security Advisory'' records as the validation set to evaluate the effectiveness of the initial set of keywords.
\item Manual optimization. We manually analy\hl{zed} the false negative results which cannot be identified by previous keywords searching and optimi\hl{z}ed the keywords to boost the detection rate on the validation set.
\item Optimization during manual investigation. We then applied the search keywords from previous step to all five GitHub repositories to scan the latent vulnerabilities. During the manual investigation of the scanning results, we further optimized the search keywords based on false positive records to minimize the false detection rate.
\end{itemize}

The final keywords are listed in Table~\ref{tab_regex}. Note that in some projects, such as Caffe and Keras, there is no expressive vulnerability update information. To resolve this problem, a comprehensive \hl{set of} latent vulnerability detection keywords is built and optimized accordingly, to scan GitHub resources. To date, all commits, issues, and pull requests of the five DL frameworks that are created on or before Jan\hl{uary} 10, 2023 are obtained with GitHub APIs, which are further input to our initial taxonomy analysis of vulnerabilities in \hl{DL} systems. \hl{The} initial statistics of the studied \hl{DL} models are reported in Table~\ref{tab_statistics-1}.


\noindent \textbf{Screening.~}
Following selection criteria are defined for dataset collection: \one~Mining established database - the vulnerability is reported in two major websites, including GitHub security advisory~\cite{githubsecurity2022} and National Vulnerability Database~\cite{NVD2022}; \two~Mining actively updated repositories - the vulnerability could be identified and fixed prior to be reported to NVD, which normally takes weeks or months for verification. Since the data in GitHub repositories could be noisy, several sub-rules are defined for preprocessing:

\begin{itemize}[leftmargin=*]
\item We only collect merged PR where all changes are approved and accepted. Therefore, such discussion and detailed information will be available and are helpful for vulnerability classification and understanding.
\item We collect PRs that match the regex rules in Table~\ref{tab_regex}.We generate a comprehensive set of regular expression rules based on the search keywords obtained through manual optimization (refer to the four-step approach).
\item There is an available patch to fix the identified security-related commits. The patch normally contains developers’ comments and the code for fixing, allowing to understand the root cause of vulnerability easily, providing information of code changes for the fixing patterns and efforts.
\end{itemize}

\begin{table*}[t]
\caption{The regular expression for vulnerability search}\label{tab_regex}
\centering
\resizebox{\linewidth}{!}{
\begin{tabular}{lp{8cm}p{9cm}}
\toprule 
\textbf{Vulnerability Type} & \textbf{Keywords} & \textbf{Regular Expression} \\
\midrule
Common & Security, Safety, Vulnerability, Threat, Violate, Malicious & (in)?secur(e$\mid$ity)$\mid$unsafe$\mid$safer$\mid$vuln$\mid$threat$\mid$violate$\mid$malicious \\
Symptom & Abort, Core Dump, Crash, Fatal, Segmentation Fault, Check Fail, Undefined Behavior & abort$\mid$core.dump$\mid$crash$\mid$fatal$\mid$seg(ment)?(ation)?(.)?fault$\mid$check.fail$\mid$fail(ing$\mid$-ed)?(.)?check$\mid$assertion.fail$\mid$undefined.behavio(u)?r$\mid$misbehave \\
Attack Method & Attack, Code Injection, Poison, Remote Code Execution, Arbitrary Code Execution, Denial of Service & attack$\mid$code.injection$\mid$poison$\mid$(remote$\mid$arbitrary).code.execution$\mid$\textbackslash brce\textbackslash b$\mid$-denial.of.service$\mid$\textbackslash bdos\textbackslash b \\
Vulnerability Database & CWE, CVE & \textbackslash bcve\textbackslash b$\mid$\textbackslash bcwe\textbackslash b \\
Permission & Authentication, Unauthorised Access, Permission & unauthenticated$\mid$gain.access$\mid$permission \\
Numeric & Divide by Zero, Integer Overflow, Integer Underflow, Integer Truncation, Floating Point Exception &  div(i(de$\mid$sion)(s)?)?.by.(zero$\mid$0)$\mid$int(eger)?.(overflow$\mid$underflow)$\mid$int(eger)?.-truncation$\mid$floating.point.exception$\mid$\textbackslash bfpe\textbackslash b \\
Memory & Out of Memory, Memory Leak, Memory Corruption, Invalid Memory Access, Arbitrary Memory Access, Heap Overflow, Stack Overflow, Out-of-Bounds (Access/Read/Write), Overflow, Overrun & out(side)?.of.memory$\mid$\textbackslash boom\textbackslash b$\mid$(data$\mid$memory).(leak$\mid$corruption)$\mid$invalid.-memory.access$\mid$access.(\textbackslash w+.)?invalid.memory$\mid$write.(\textbackslash w+.)?immutable.-memory$\mid$access.(\textbackslash w+.)?undefined.memory$\mid$arbitrary.memory.(read$\mid$write-$\mid$access)$\mid$heap.(buffer.)?overflow$\mid$stack.overflow$\mid$out(side)?.of.bound(s)?$\mid$-\textbackslash boob\textbackslash b$\mid$overflow$\mid$overrun \\
Resource Management & Use After Free, Use of Uninitialized Resource & use.after.free$\mid$uninitiali[s$\mid$z]ed \\
Pointer & Null Pointer Dereference, Null Pointer Exception, Reference Binding to Null Pointer & null(.)?(pointer$\mid$ptr)?(.)?(deref(erence)?$\mid$exception)$\mid$deref(erenc(e$\mid$ing))?.of.-null(.)?(pointer$\mid$ptr)?$\mid$\textbackslash bnpe\textbackslash b$\mid$reference.binding.to.null(.)?(pointer$\mid$ptr)? \\
Data Validation & Validation & \textbackslash bvalidat(e$\mid$ion)(d$\mid$s)?\textbackslash b \\
Other & Dead Code, Race Condition, Format String, Type Confusion, Shape Check & dead(.)?code$\mid$\textbackslash brac(e$\mid$ing)$\mid$thread.safe$\mid$format.string$\mid$string.format$\mid$type.-confusion$\mid$shape.check$\mid$check.shape \\
\bottomrule
\end{tabular}
}
\end{table*}


\noindent \textbf{Implementation.~}
Specifically, to reduce the data harvest effort, we implement a data acquisition system to automate the data collection process, as illustrated in Figure~\ref{fig:data_collection}. \hl{This process involves five datasets: \textit{NVD}, \textit{GitHub}, \textit{Raw}, \textit{Distilled}, and \textit{Manual}. The \textit{NVD} dataset houses official CVE records, offering professional vulnerability descriptions. The \textit{GitHub} dataset comprises code, commits, pull requests, and issues from DL projects. The \textit{Raw} dataset collects unprocessed data directly crawled from GitHub, while the \textit{Distilled} dataset retains data that have been automatically filtered or processed. Finally, the \textit{Manual} dataset is dedicated to data that have been processed manually.}

For public vulnerability data, we implement a \textit{NVD Crawler} to query the CVE records by DL framework name from NVD. In addition, for TensorFlow, which includes well-maintained security advisories, we implement \textit{TFSecurityAdvisory Parser} to extract useful information from security advisory files, such as title, CVE Number, patches (the patching approaches), issue description, impact (the symptom or consequence), vulnerable versions, mitigation, credits, attribution. These data will be used as a reference to improve the performance of our latent vulnerability detection keywords.

For latent vulnerability data, we clone the five repositories \hl{on} the local machine and execute \textit{GH Filter}, which integrates with \hl{the} GitHub CLI command to extract all pull requests, commits and issues. \hl{Only a} simplified version of the logs is returned. We then use \textit{GHCrawler (PR)} to query \hl{the} GitHub REST API and harvest completed pull request information. 
Then, we applied a latent vulnerabilit\hl{y} search on the pull request records to expose those vulnerability patches. Finally, \textit{GHCrawler (Commit)} will crawl the completed commit information for \hl{the} corresponding vulnerability patches.

\subsection{Classification and Labeling Process}
\label{sec_classification_labelling}
In order to effectively characterize the vulnerabilities in \hl{DL} systems, we focus on three distinguished aspects, \ie the root causes, the identified symptoms, and the patching process of vulnerabilities, by following the widely adopted open coding procedure~\cite{holton2007coding}. 
The data is specifically curated with `impact' and `patches' information.
Meanwhile, we performed an iterative manual labeling process involving the authors' efforts to substantially investigate the identified latent vulnerabilities with their expertise and development knowledge.
The iterations are as follows:

\noindent \textbf{\textit{Iteration 1.}} \hl{First we} create the initial search pattern following ~\cite{bosu2014identifying, zhou2017automated}. Official vulnerabilities are used to improve the detection performance of search keywords. The search patterns are iteratively improved and achieved the best detection results (\eg false negatives and positives). Optimal search patterns are applied to the three sources (\ie PRs, commits, and issues) of data for latent vulnerability detection. 
Specifically, we apply latent search for PR, and then according to the vulnerable PR results, find the corresponding commits and Issues. Within this iteration, an initial version of classification and labeling strategy is obtained, according to the search patterns and a preliminary discussion. 



\begin{table}[t]
\caption{Latent vulnerabilities identified in Iteration 1.}\label{tab_statistics_2}
\centering
\resizebox{0.8\linewidth}{!}{
\begin{tabular}{lrrr} 
\toprule
\textbf{Framework} & \textbf{NVD Data} & \textbf{Pull Requests} & \textbf{Commits} \\ 
\midrule
TensorFlow & 407 & 1,950 & 3,072 \\ 
PyTorch    & 1   & 286   & 930   \\ 
OpenCV     & 35  & 735   & 2,137 \\ 
Keras      & 0   & 97    & 373   \\ 
Caffe      & 0   & 96    & 241   \\ 
\bottomrule
\end{tabular}
}
\end{table}

\noindent \textbf{\textit{Iteration 2.}} For detection results from Iteration 1, two authors manually review PRs, commits, and issues, including source code, title, description, comments, and discussion. After review, authors independently label records from seven aspects according to the review of CWE List 4.9 from the research concept: (1) taxonomy, (2) CWE ID, (3) root cause, (4) fixing pattern, (5) symptoms, and whether it is a (6) vulnerability and (7) DL-related. As this paper mainly focuses on DL-related vulnerability, the taxonomy was further broken down into DL-specific taxonomy. The results were then compared, and the two authors have demonstrated an almost perfect agreement by achieving a Kappa coefficient of 0.895~\cite{banerjee1999beyond}. For the disagreements in the labels, a discussion was conducted to revise the classification and labeling strategy accordingly.

\noindent \textbf{\textit{Iteration 3.}} Two authors thoroughly reviewed all data with the revised strategy again, and had a comparative discussion of the outcome, including all disagreements regarding all seven aspects. Eventually, a final agreement on the taxonomy of the root causes is reached by resolving all disagreements.

\subsection{Research Questions}
To ensure \hl{that} the study comprehensively analyzes the identified vulnerabilities, we answer \hl{the} following research questions that are concerned with the root causes of vulnerabilities in DL systems and challenges for patching the vulnerabilities:

\begin{itemize}[leftmargin=9mm]
\item[\textbf{RQ1:}] What are the common root causes of \hl{DL}-specific vulnerabilities in the \hl{DL} frameworks?
\item[\textbf{RQ2:}] What are the challenges in detecting vulnerabilities?
\item[\textbf{RQ3:}] In terms of fixing the vulnerabilities, what are the main challenges, and how to address them?
\end{itemize}

It is expected to understand how vulnerabilities in \hl{DL} systems arise by studying RQ1. The summary of common root causes will support the answers to RQ2 and RQ3 \hl{by} providing more accurate explanations. Moreover, new challenges imposed by the key differences between \hl{DL} systems and traditional software are investigated. 

\section{Experimental Results}
\label{sec:experimental_results}
In this section, we present the experimental results. 
We have reviewed the latest CWE List report and categorize the vulnerability for its most relevant software weakness types, which is illustrated in Table~\ref{tab_cwe_vuln}. Overall, 
\typesOfVuln different types of vulnerabilities from CWE List 4.9 are identified. However, still near 9\% of the vulnerabilities could not be well aligned with the current CWE frameworks. 


\begin{table*}[t]
\caption{The vulnerability types based on CWE version 4.9 from analytic data}\label{tab_cwe_vuln}
\centering
\resizebox{0.9\linewidth}{!}{
\begin{tabular}{@{}lllc@{}}
\toprule 
\textbf{Pillar ID} & \textbf{CWE Pillar} & \textbf{CWE Class/Base in Deep Learning Systems}  &  \textbf{Proportion}  \\
\midrule
CWE-664 & Improper Control of a Resource Through its Lifetime  
        & CWE-22, CWE-94, CWE-118, CWE-119, CWE-120, CWE-125, ...
        & 33/418 \\
CWE-682 & Incorrect Calculation  
        & CWE-131, CWE-190, CWE-191, CWE-193, CWE-369, CWE-1339
        & 6/13 \\
CWE-691 & Insufficient Control Flow Management 
        & CWE-248, CWE-362, CWE-431, CWE-617, CWE-670, CWE-674,...
        & 9/94\\
CWE-693 & Protection Mechanism Failure 
        & CWE-354, CWE-778, CWE-311
        & 3/101 \\
CWE-703 & Improper Check or Handling of Exceptional Conditions  
        & CWE-237, CWE-241, CWE-252, CWE-280, CWE-754, CWE-755
        & 6/59\\
CWE-707 & Improper Neutralization  
        & CWE-20, CWE-77, CWE-1284, CWE-1285, CWE-1287 
        & 5/149\\
CWE-710 & Improper Adherence to Coding Standards  
        & CWE-476, CWE-561, CWE-1357
        & 3/198\\
\bottomrule
\end{tabular}
}
\end{table*}

\subsection{RQ1: Root Causes of Vulnerabilities}
\label{sec:root_causes}
As an immediate outcome of the classification and labeling process from Section~\ref{sec_classification_labelling}, the root cause of the  \vulnUndreReview vulnerabilities in DL systems are categorized into 7 groups, as detailed below. 

\begin{figure}[t]
\centering
\includegraphics[width=\linewidth]{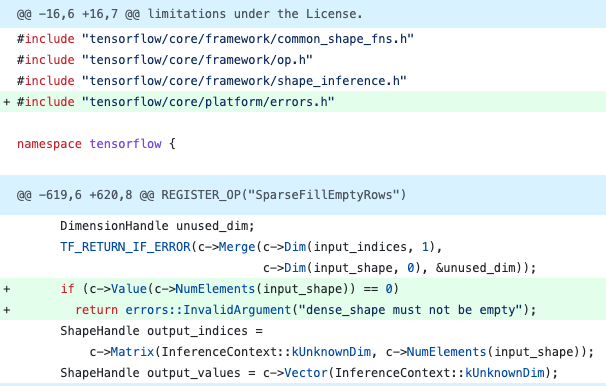}
\caption{Example of CVE-2021-37676 for resource control.}
\label{fig:root_cause_1_1}
\end{figure}

\noindent \textbf{Root cause 1: Insufficient computational resource control.~}
\label{subsubsection:root_cause_1}
A sufficient control level of resources is not maintained throughout the development lifetime.
In \hl{DL} systems, computational resources are substantial, which requires the software \hl{to} strictly follow the instructions to create, use and destroy the resources, including \hl{the} tensor, memory, and pointer. Among the 950 resource control related vulnerabilities, the subjects are diverse and the operations are different, making the developers stuck in a difficult situation and hard to handle the resources correctly. We have identified the following vulnerability subsets of root cause 1:

\begin{itemize}[leftmargin=*]    
\item Improper resource validation. Due to excessive usage of high-performance computational resources, developers need to properly validate the incorporated resources and the \hl{elements defined} in the code, including GPU availability, algorithm parameters, tensor shape, rank, and dimension. \hl{A} classical issue related to tensor shape is found in CVE-2021-37676 in Figure~\ref{fig:root_cause_1_1}, in which \hl{inference of} tensor shape does not validate the input, leading to attacks exploiting unexpected privileges to launch undefined model behavior.
\item Incorrect resource usage. The developers need a proper definition of the operation to the resources, in which of the DL-enabled systems the major concerns are from \hl{the} data and memory buffer. Handling data type differences and conversion could be an even worse problem when dealing with GPU kernel. \hl{A} heap buffer overflow vulnerability is CVE-2019-16778 in TensorFlow, which represents a typical example of incorrect type conversion from \texttt{int64} to \texttt{int32}. Similarly to \hl{the} out-of-bounds memory access issue, operations within the memory buffer bounds require proper restriction to prevent memory overflow, corruption, and invalid array write. \hl{In particular}, the tensor is a key element in \hl{DL}. Thus, giving explicit instructions on resource usage is critical.
\item Uncontrolled resource release. The excessive range of resources for \hl{DL} have brought developers a new landscape of managing resource after use. In addition to the weakness of `Double Free', `Use After Free' becomes common for the data input usage, such as the malicious behavior when decoding PNG images in CVE-2022-23584.
\end{itemize}

\bigbreak
\colorbox{gray!20}{
\begin{minipage}{0.43\textwidth}
\protect\hl{Improper resource validation, incorrect resource usage, and uncontrolled resource release are the root causes of vulnerabilities due to insufficient computational resource control.}
\end{minipage}
}
\bigbreak

\begin{figure}[t]
\centering
\includegraphics[width=\columnwidth]{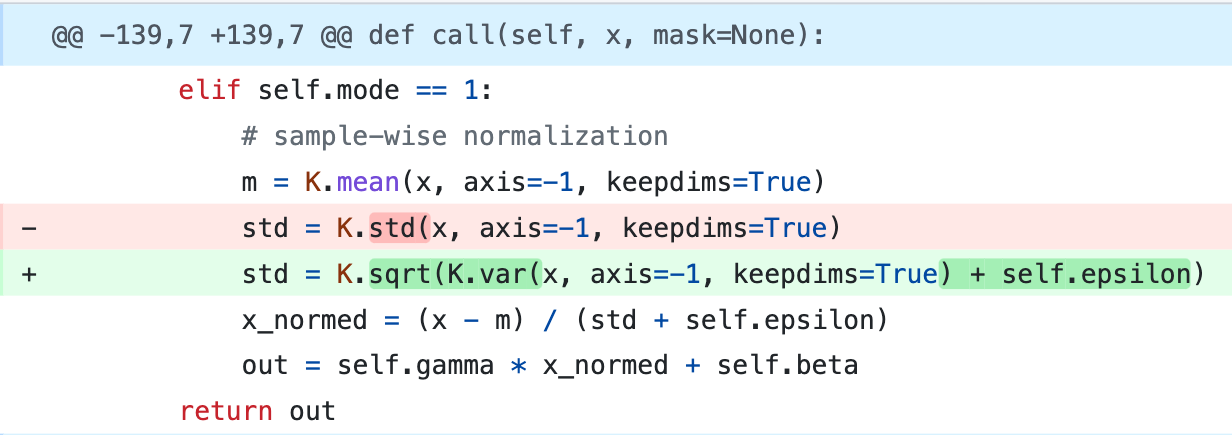}
\caption{An example of fixing incorrect gradient computation in Keras issue \#2960.}
\label{fig:root_cause_2_1}
\end{figure}

\noindent \textbf{Root cause 2: Incorrect security-critical gradient calculation\&operation.~}
\label{subsubsection:root_cause_2} 
If security-critical calculation is not \hl{performed correctly}, it could potentially lead to incorrect resources allocation and incorrect privileges assignment. In DL, massive gradient calculations are required for the learning and inference capability. Since the calculation is sensitive to both training and testing stages, developers are prone to ignore the issues, which will sometimes be hard to fix. Two major types are among the 418 calculation caused vulnerabilities:

\begin{itemize}[leftmargin=*]
\item Wrong calculation. Developers may ignore the validity of the calculation throughout \hl{DL} development, \hl{leading} to vulnerabilities such as `Divide by Zero', `Incorrect Calculation', and `Incorrect Calculation of Buffer Size', according to CWE List version 4.9. \hl{Gradient computation is common} in \hl{DL} since the standard deviation computation can sometimes be numerically unstable. In Keras issue \#2960, the output is rewritten to avoid a division by zero in \texttt{BatchNormalization} layer. Figure~\ref{fig:root_cause_2_1} demonstrates the fix \hl{for} the issue.
\item Integer underflow/overflow. Incrementally adding or subtracting a value from another may cause integer overflow or underflow, respectively. In OpenCV, \hl{there} can be serious security issues of remote code execution or denial of service when loading images remotely, such as CVE-2017-12864. 
\end{itemize}

\bigbreak
\colorbox{gray!20}{
\begin{minipage}{0.43\textwidth}
\protect\hl{Wrong calculation, and integer underflow/overflow are the root causes of vulnerabilities due to incorrect security-critical gradient calculation\&operation.}
\end{minipage}
}
\bigbreak

\noindent \textbf{Root cause 3: Insufficient control flow management in parallel computation.~}
\label{subsubsection:root_cause_3}
There is \hl{no} dedicated code implementation \hl{that provides} sufficient control flow management during execution, resulting in security weakness to alter the control flow. Developing DL systems often involves parallel computing using shared GPU, threads, and memory. \hl{Compared} to traditional software systems, DL systems are more prone to control flow management vulnerabilities. \hl{In total}, 112 vulnerabilities are identified and categorized into two subsets:

\begin{itemize}[leftmargin=*]
\item Unexpected behavior. To better support parallel computing for \hl{DL}, developers need to consider the race condition carefully, which could occur in the GPU fill function (\eg PR\#10298 in TensorFlow~\cite{tf2017pr10298}) and model generators (\eg PR\#5049 in Keras~\cite{keras2017pr5049}). Such unexpected behaviors will cause the output to be non-deter\-ministic as well as crash with \texttt{ValueError}. To fix, developers can either fix the control flow or synchronization, or add dedicated management checkers for the index.
\item Missing handling exceptions. Since exceptions are generally an opportunity for the run-time test, the developers will add handlers to catch and handle the exceptions. For \hl{DL}, there will be many defined blocks \hl{that} implement functions, such as OpenMP parallel blocks. A typical example is raised in PR\#4857 in PyTorch~\cite{pytorch2018pr4857}, which \hl{does not} allow throwing exceptions across OpenMP constructs.
\end{itemize}

\bigbreak
\colorbox{gray!20}{
\begin{minipage}{0.43\textwidth}
\protect\hl{Unexpected behavior, and missing handling exceptions are the root causes of vulnerabilities due to insufficient control flow management in parallel computation.}
\end{minipage}
}
\bigbreak


\noindent \textbf{Root cause 4: Lack of protection mechanism.~}
\label{subsubsection:root_cause_4} 
For adversarial attacks, the root cause of such vulnerabilities is due to unable to provide sufficient defense. A `missing' protection mechanism is particularly prevalent in \hl{DL}. One typical example is GPU operation that reads invalid memory, causing the `Insufficient Verification of Data Authenticity' in CVE-2021-41203. Providing Checkingpoints loading the infrastructure is demanded.

\bigbreak
\colorbox{gray!20}{
\begin{minipage}{0.43\textwidth}
\protect\hl{Lack of protection mechanism is a prevalent  vulnerability in DL.}
\end{minipage}
}
\bigbreak

\begin{figure}[t]
    \centering
    \includegraphics[width=\columnwidth]{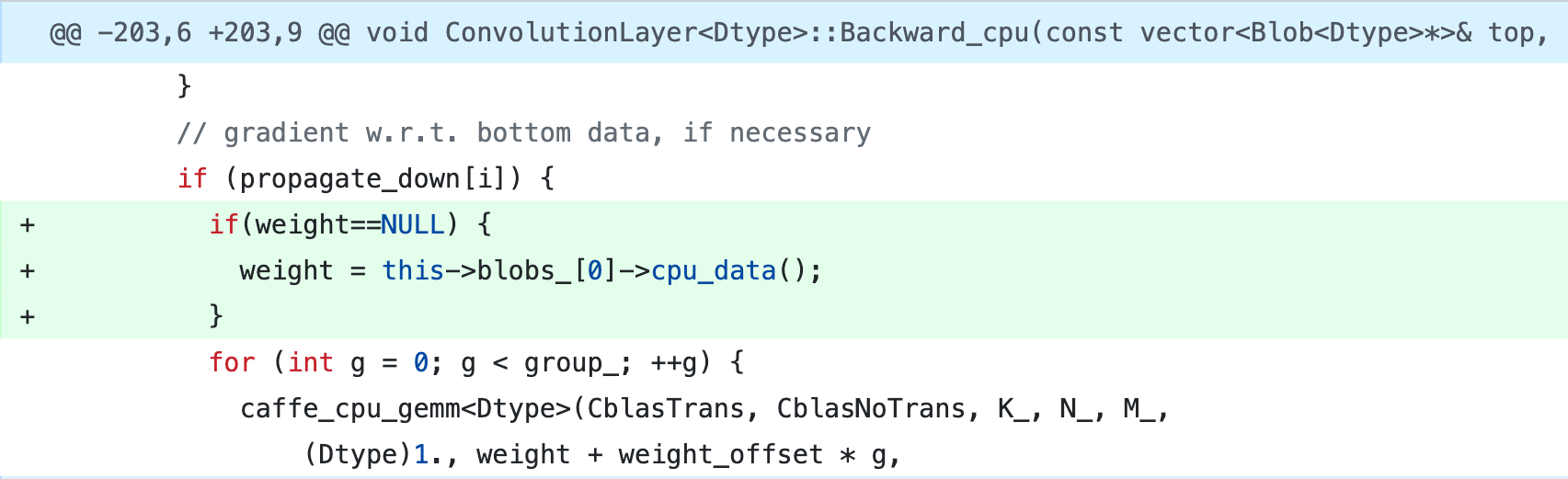}
    \caption{Example of PR\#1048 for exceptional conditions.}
    \label{fig:root_cause_5_1}
\end{figure}

\noindent \textbf{Root cause 5: Improper DL corner case handling.~}
\label{subsubsection:root_cause_5}
During normal operation, it is desired to consider handling exceptional conditions. \hl{DL} encapsulates more input and output variables, return value, and run-time model state. Thus, how to deal with more corner cases is \hl{difficult} for \hl{DL} systems. In PR\#7071 in Keras~\cite{keras2018pr7071}, a simple but effective example is related to the values of the data label. The patch adds one more condition to check the label values. Figure~\ref{fig:root_cause_5_1} shows PR\#1048 in Caffe~\cite{caffe2014pr1048} where a code snippet fixes the exceptional condition of setting the weight pointer as \texttt{NULL} for convolutional layer training. For exceptional conditions, as we can see from the code snippets, PR\#1048~\cite{caffe2014pr1048} and PR\#7071~\cite{keras2018pr7071} have no common variables, and vulnerabilities are hard to identify if \hl{they do} not presenting any understandings of DL.

\bigbreak
\colorbox{gray!20}{
\begin{minipage}{0.43\textwidth}
\protect\hl{Improper DL corner case handling is one of the root cause of DL vulnerabilities due to the challenge compounded by the complexity of DL processes and the need for specialized DL knowledge.}
\end{minipage}
}
\bigbreak

\begin{figure}[t]
    \centering
    \includegraphics[width=\columnwidth]{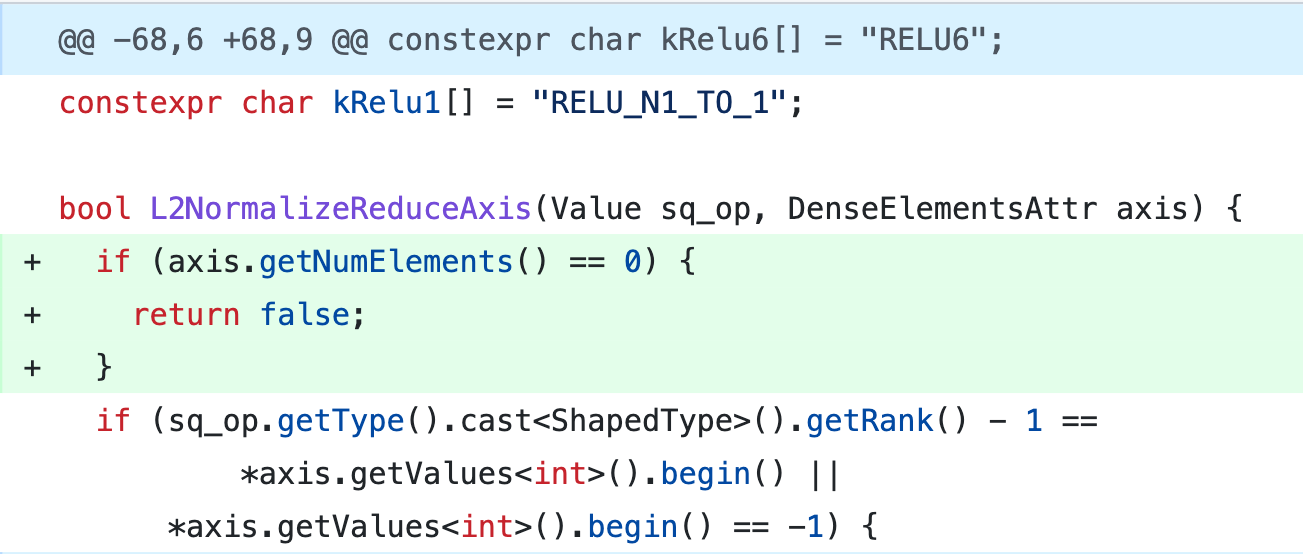}
    \caption{Example of CVE-2021-37689 for coding rules.}
    \label{fig:root_cause_7_1}
\end{figure}

\noindent \textbf{Root cause 6: Input/output neutralization.}
\label{subsubsection:root_cause_6}
The root cause of such vulnerability is being unable to guarantee the message flowed as input and output in the system. Our dataset dominantly originates from the improper validation of the specified type and quantity of input. Some may \hl{be} related to memory and GPU allocation, such as the PR\#38417 in TensorFlow~\cite{tf2020pr38417}. When there is no GPU in the system, the device tracer should not be created to avoid out-of-memory. It may happen in the feature extraction stage, i.e., `CVE-2016-1517' for OpenCV. The attacker can launch remote code execution on the victim's machine by injecting an infected image with OpenCV. Moreover, for DL systems, it would be much more complicated for developers to analyze the latent attack surface and check that the specified input and output are both `safe'.

\bigbreak
\colorbox{gray!20}{
\begin{minipage}{0.43\textwidth}
\protect\hl{Unable to neutralize input/output is a major root cause of vulnerability in DL.}
\end{minipage}
}
\bigbreak

\noindent \textbf{Root cause 7: Improper adherence to coding rules.~}
\label{subsubsection:root_cause_7} 
The root cause of \hl{this} vulnerability refers to not following certain coding rules of development. Development for DL is a rapidly developing area, for which a significant amount of code is implemented to deliver the functional and non-functional requirements. \hl{Until} now, given the booming programming language and framework communities for \hl{DL}, there \hl{are} no coding standards or rule\hl{s} for developers. In our dataset, `Dead code', `Null Pointer Dereference' and `Reliance on Uncontrolled Component', are three main sub-root causes. Figure~\ref{fig:root_cause_7_1} is an example of null pointer dereference caused by a crafted \texttt{TFLite} model from manipulation of the MLIR optimization function of \texttt{L2NormalizeReduceAxis} operator.

\bigbreak
\colorbox{gray!20}{
\begin{minipage}{0.43\textwidth}
\protect\hl{Dead code, null pointer dereference, reliance of uncontrolled component are the root causes of vulnerabilities due to improper adherence to coding rules.}
\end{minipage}
}
\bigbreak

\subsection{RQ2: Challenges in Vulnerabilities Detection}
\hl{To} understand RQ2, we go through the dataset focusing on two aspects, the input to trigger the vulnerabilities (its type, range, and visibility), and the output and behavior (the impact).
It \hl{should be} noted that, for developers, figuring out the exact location and \hl{performing} a specific analysis of the vulnerability in DL is challenging. However, with the developers discussion corpus and source code, 
this is by far the best way to provide us the insight from our dataset regarding the vulnerability detection challenges in \hl{DL} systems.

\noindent \textbf{Challenge 1: DL artifacts testing.}\label{subsubsection:challenge_1} 
While there is a hard-coded relationship between inputs and outputs in traditional software systems, inclusively and effectively designing a tool as \hl{a} test oracle boots the software development efficiency. However, it is still a challenging task to test the system to date. Now\hl{,} with learning and inference capability, \hl{DL} systems attempt to extrapolate the relationship from massive data as a core module in the system, such as Microsoft Outlook and OpenDNS. To achieve a specific function in the DL systems, the development team fac\hl{es} many more options ranging from data, features, algorithms, and models.
As discussed in Section~\ref{sec:root_causes}, the input and output in DL systems are highly dynamic and do not present a consistent pattern. The input could be varied on a case-by-case basis, such as an empty list (\eg PR\#58651 in PyTorch~\cite{pytorch2021pr58651}) and a specially crafted XML file input (\eg CVE-2019-5063). When input flows to a specific model layer, such as LSTM/GRU layer (\eg CVE-2020-26270), failure to access CUDA in the back-end results in a query-of-death vulnerability. 

On \hl{the other} hand, \hl{DL} artifacts are more intrinsically linked to the infrastructure, projecting the inference and decision into a hierarchical structure. Even for a securely saved model, it can trigger the `Use of Uninitialized Resource' vulnerability while building the computation graph (\eg CVE-2020-26271) or during code execution (\eg CVE-2020-26266). These model-related behaviors become broader and generate more critical impacts in terms of the DL system components and vulnerability types, respectively. It is substantially challenging for developers to determine an effective DL artifact testing protocol, which refers to the testing oracle in software. Some works have started investigat\hl{ing} issues such as tensor property~\cite{harzevili2022characterizing} and \hl{DL} operators~\cite{zhang2021duo} by expanding the capability of existing test suite to design novel security test cases. As a significant outcome of the DL systems vulnerability taxonomy, future research may target a general test oracle problem of DL systems testing and achieve an automated solution to deliver the testing process. DL experts can play a pivotal role in testing design work and provide important definition criteria.

\bigbreak
\colorbox{gray!20}{
\begin{minipage}{0.43\textwidth}
\protect{\hl{The dynamic input-output relationship and intrinsically link with underlying infrastructure make it challenges to test DL artifacts.}}
\end{minipage}
}
\bigbreak

\noindent \textbf{Challenge 2: Hardware support.}\label{subsubsection:challenge_2} 
Currently, general and specific hardware for DL system has been complicated since it generates the data in a different form, such as the sensor data, signal input, middle-ware data flow, and output. \hl{Hardware support is observed to be} less likely to be detected, such as CUDA development kits for GPU support (\eg PR\#8246 in PyTorch~\cite{pytorch2018pr8246} and PR\#633 in Caffe~\cite{caffe2014pr633}). 

Thus, a developer who has less experience with the hardware will be in a disadvantageous position when implementing \hl{DL} systems. In PR\#5376 in PyTorch~\cite{pytorch2018pr5376} and PR\#42615 in TensorFlow~\cite{tf2020pr42615}, the developers triggered \hl{runtime} errors without correct hardware settings and configurations. Specifically, in a multi-GPU setup, PyTorch will attempt to utilize all GPU hardware for the best performance. However, this behavior may randomly crash with an \textit{Out-of-Memory} error because of an imbalance of computing capability among GPUs. Unlike PyTorch, TensorFlow is free from such vulnerability because it can automatically discard those GPUs without sufficient computing capabilities. However, a callback in \texttt{DynamicPartitionOpGPU} will run outside the correct GPU context and eventually crash. Although PyTorch can be configured to avoid this problem, the inaccuracy or insufficient error message points developers in a wrong debug direction, making it challenging to recognize the vulnerabilities.

\bigbreak
\colorbox{gray!20}{
\begin{minipage}{0.43\textwidth}
\protect\hl{The stealthy nature of vulnerabilities linked to hardware support in DL system complicates the vulnerability detection.}
\end{minipage}
}
\bigbreak

\noindent \textbf{Challenge 3: Domain knowledge.}\label{subsubsection:challenge_3}
Unlike traditional attacking behavior, one main goal of attacking a \hl{DL} system is to introduce malicious data and manipulate the model behavior, misleading the prediction, clustering or regression output. The overall \hl{DL} lifecycle involves a dynamic process of processing raw data, feature representation, model training, and evaluation, in which the logic related (see root cause \hyperref[subsubsection:root_cause_1]{1}, \hyperref[subsubsection:root_cause_1]{3}, and \hyperref[subsubsection:root_cause_1]{5} in \S~\ref{sec:root_causes})
and artifact related (see
root cause \hyperref[subsubsection:root_cause_1]{2}, \hyperref[subsubsection:root_cause_1]{6}, and \hyperref[subsubsection:root_cause_1]{7} in \S~\ref{sec:root_causes}
vulnerabilities are difficult to discover using existing techniques.
One typical example is from PR\#5161 in OpenCV~\cite{opencv2019pr5161}, in which the developer has limited knowledge about convergence algorithms. Insufficient knowledge of the algorithm in this issue leads to incorrect implementation and undefined behavior. This type of vulnerability is hard to detect because developers normally assume that the implementation of \hl{the} algorithm is correct. As a widely applied \hl{DL} algorithm, this vulnerability can cause a wide range of impacts.

\bigbreak
\colorbox{gray!20}{
\begin{minipage}{0.43\textwidth}
\protect\hl{Addressing the challenge of detecting logic-related and artifact-related vulnerabilities requires a strong foundation in domain knowledges.}
\end{minipage}
}
\bigbreak

\noindent \textbf{Challenge 4: Third-party library support.}
\label{subsubsection:challenge_4}
Like other software, \hl{DL} systems \hl{relied heavily} on third-party libraries. \hl{Although} these libraries can reduce \hl{the} coding effort during development, \hl{they} introduce significant security risks. Insufficient understanding of the third-party code might hinder the vulnerability detection, thus increasing the patching effort.

It refers to the challenge of legitimating the third-party library in use and installing well\hl{-}supported in place. In PR\#13692 in OpenCV~\cite{opencv2019pr13692}, developers discover a vulnerability related to third-party library compatibility when aim\hl{ing} to enable an \hl{asynchronous external} API in OpenCV. However, the support of FP32 networks is missing on Intel's Myriad plugin. The first patch actually introduces a check for the type of inference engine to neutralize the potential crash. However, this approach dramatically increases the execution time even generat\hl{ing} incorrect results due to running unnecessary tests. After digging into the details of the external API, a new patch is done to eliminate this issue by refactoring the execution flow to avoid supernumerary assertion\hl{s}. To fix the lack of support issues, developers have to patch the vulnerability twice.

\bigbreak
\colorbox{gray!20}{
\begin{minipage}{0.43\textwidth}
\protect\hl{Using third-party libraries in developing DL systems without sufficient understanding of the code will significantly increase security risks, presenting challenges for detecting and preventing vulnerabilities.}
\end{minipage}
}
\bigbreak

\subsection{RQ3: Challenges in Vulnerability Patching}

To answer RQ3, we carefully reviewed the available patch and the developer\hl{s'} discussion in comments to understand the challenges in fixing the \hl{DL} systems vulnerabilities. In the meantime, the respective efforts are conducted by summarizing the line and file changes, discussion comments, and time stamps. Overall, the goal is to answer RQ3 from two different perspectives: one is from the difficulty level of the vulnerability patch, and \hl{the other} is the response time to close the fix loop. Complete \hl{statistical} results are included in our replication package~\cite{package2022investigation}.
In this section, we outline the identified challenges and major findings and propose our supplementary solution to extend the actionable implications from the basis of \hl{the} CWE List framework.

\noindent \textbf{Challenge 5: Vulnerability localisation.}
\label{subsubsection:challenge_5}
In the core of CWE List, it conveys a comprehensive list of individual CWE weaknesses for a range of products, such as J2EE, ASP.NET, UNIX, and Windows. It also provides developers \hl{with} guidelines to facilitate the vulnerability localization and fixing. However, it remains unclear with the current CWE List for developers to deal with the vulnerabilities in the \hl{DL} context. Particularly when the vulnerability is related to either logic or artifacts aspect\hl{s}, it is difficult for developers to determine every value and status for every stage during the lifecycle, while the identical relationship between input and output is a typical way to localize the vulnerability. Particularly when some additional features are included to provide benefits for high-performance machine learning model execution, a simple alteration to invalidate the assumptions could cause significant security issues (\eg CVE-2022-23594 for TFG dialect of TensorFlow, and CVE-2022-23588 for Grappler optimizer). 

\bigbreak
\colorbox{gray!20}{
\begin{minipage}{0.43\textwidth}
\protect\hl{The current CWE List's guidance on addressing vulnerabilities in the DL context remains unclear, posing challenges for developers in localizing vulnerabilities using standard approaches.}
\end{minipage}
}
\bigbreak

\noindent \textbf{Challenge 6: Patching process.}
\label{subsubsection:challenge_6}
Since it has been difficult for developers to efficiently identify the vulnerability in the collected patches, we observed the following actions conducted by the developer community. Developers are now facing much more challenges in the new DL system landscape. Normally, it includes the actions of vulnerability reproducing, unit testing, and community involvement during the patching process.
\begin{itemize}[leftmargin=*] 
\item Unit \hl{t}est: It is still a direct practice in the patch process. In PR\#24674~\cite{tf2019pr24674} and PR\#5349~\cite{tf2016pr5349}, the developers patched the vulnerabilities along with the corresponding unit test. Unit testing ensures the quality of the patch and is easy to debug. \hl{In addition}, by reviewing the unit test, the reviewers can quickly understand the context of the vulnerability and its impact. However, the tasks of adding positive test cases and complete unit tests \hl{are non-trivial}, such as CVE-2022-23559 for integer overflow in TFLite. 

\item Vulnerability \hl{r}eproduction: This is seldom observed in our dataset though it is a general approach for patching. One typical example following this method is in PR\#36856~\cite{tf2020pr36856} and Issue\#36692~\cite{tf2020pr36692}. The developer recognised a segmentation fault due to improper padding for convolution operation. The vulnerability was reported in detail with error logs and codes for vulnerability reproducing. However, only a limited portion of patches has followed this practice. It challenges the reviewers to countenance the patch in a timely manner.

\item Community \hl{i}nvolvement: Providing sufficient community support for the patch process is critical for vulnerabilities in DL systems. In PR\#5585~\cite{pytorch2018pr5585}, the developer patches a memory leak vulnerability due to improper release of file resources in a multiprocessing case. During the patch process, a few communities reviewers are involved to raise the questions and pointed out the patching issues, including typos, potential problems, and possible optimization. One of the reviewers also questions the necessity of the patch. Positive involvements of the community can force the developer to get better understanding of the vulnerability and improve the patch quality despite the extension of patch time.
\end{itemize}

\bigbreak
\colorbox{gray!20}{
\begin{minipage}{0.43\textwidth}
\protect\hl{Ensuring robust unit testing, accurate vulnerability reproduction, and active community involvement presents a significant challenge in the effective identification and resolution of DL vulnerabilities.}
\end{minipage}
}
\bigbreak



\begin{figure}[t]
\centering
\includegraphics[width=\linewidth]{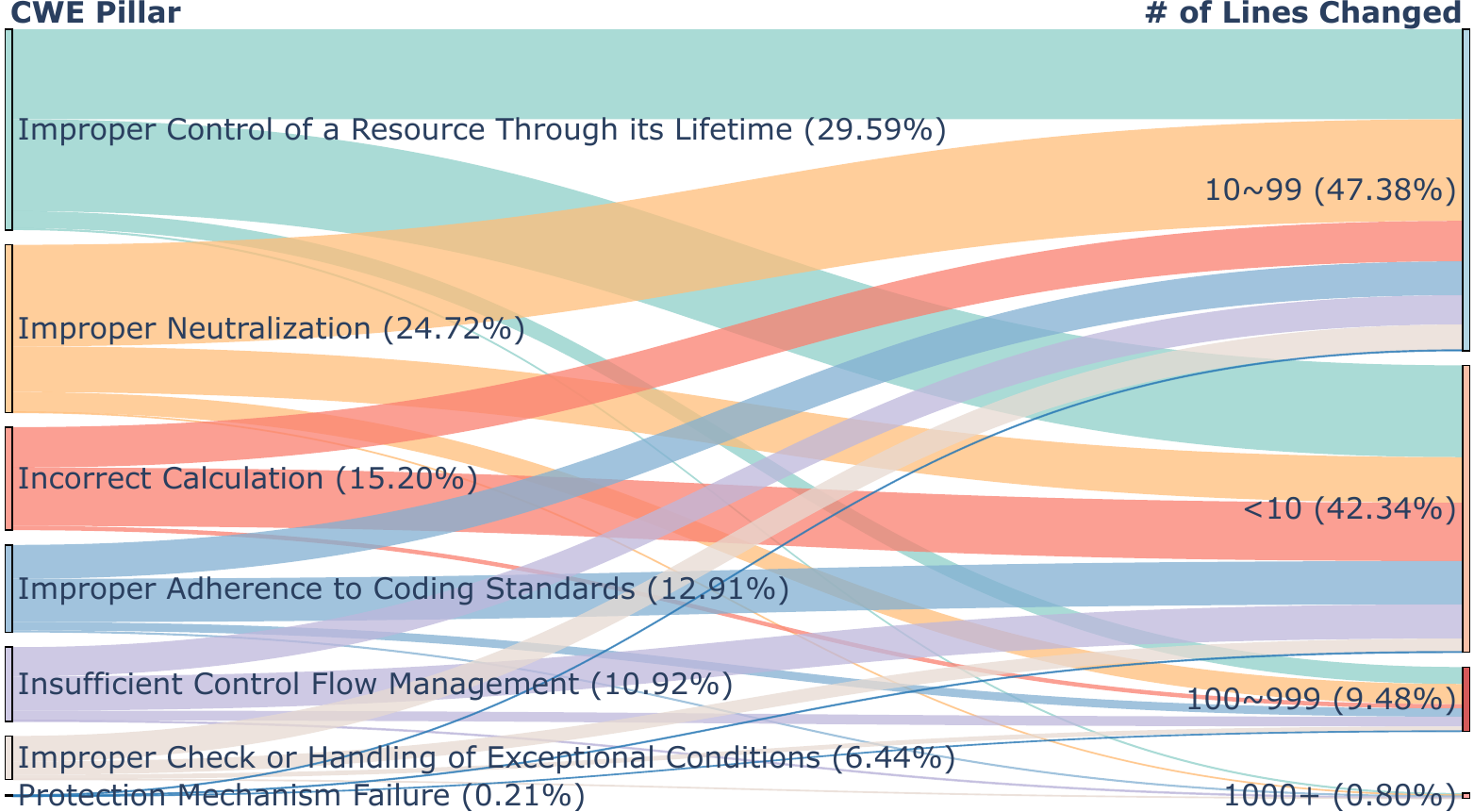}
\caption{CWE pillar to line changed mapping.}
\label{fig:pillar_to_line_changed}
\end{figure}

\begin{figure}[t]
\centering
\includegraphics[width=\linewidth]{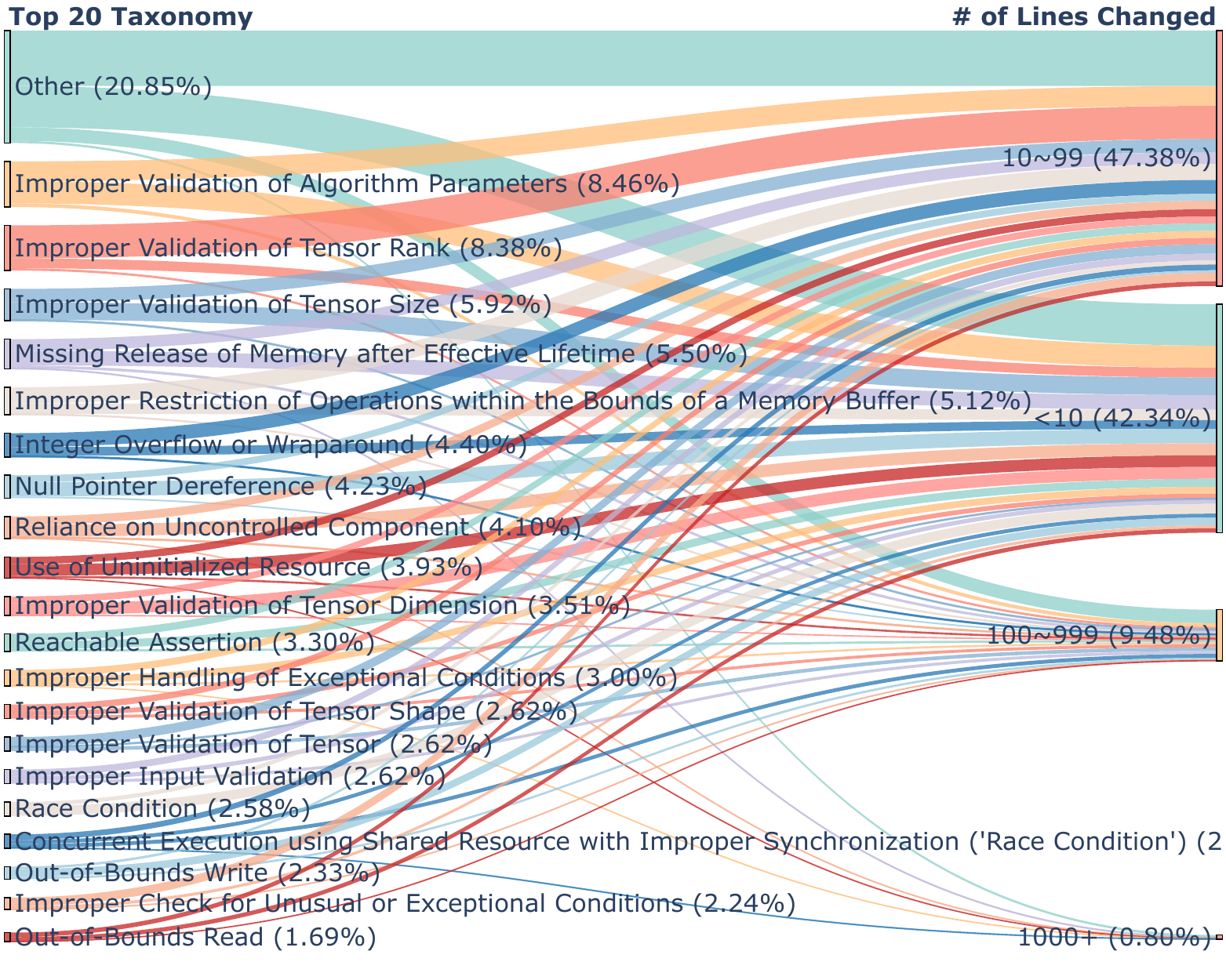}
\caption{Top 20 taxonomy to line changed mapping.}
\label{fig:taxonomy_to_line_changed}
\end{figure}


\begin{figure*}[t]
\centering
\includegraphics[width=\linewidth, trim=1cm 1.5cm 1cm 1cm, clip]{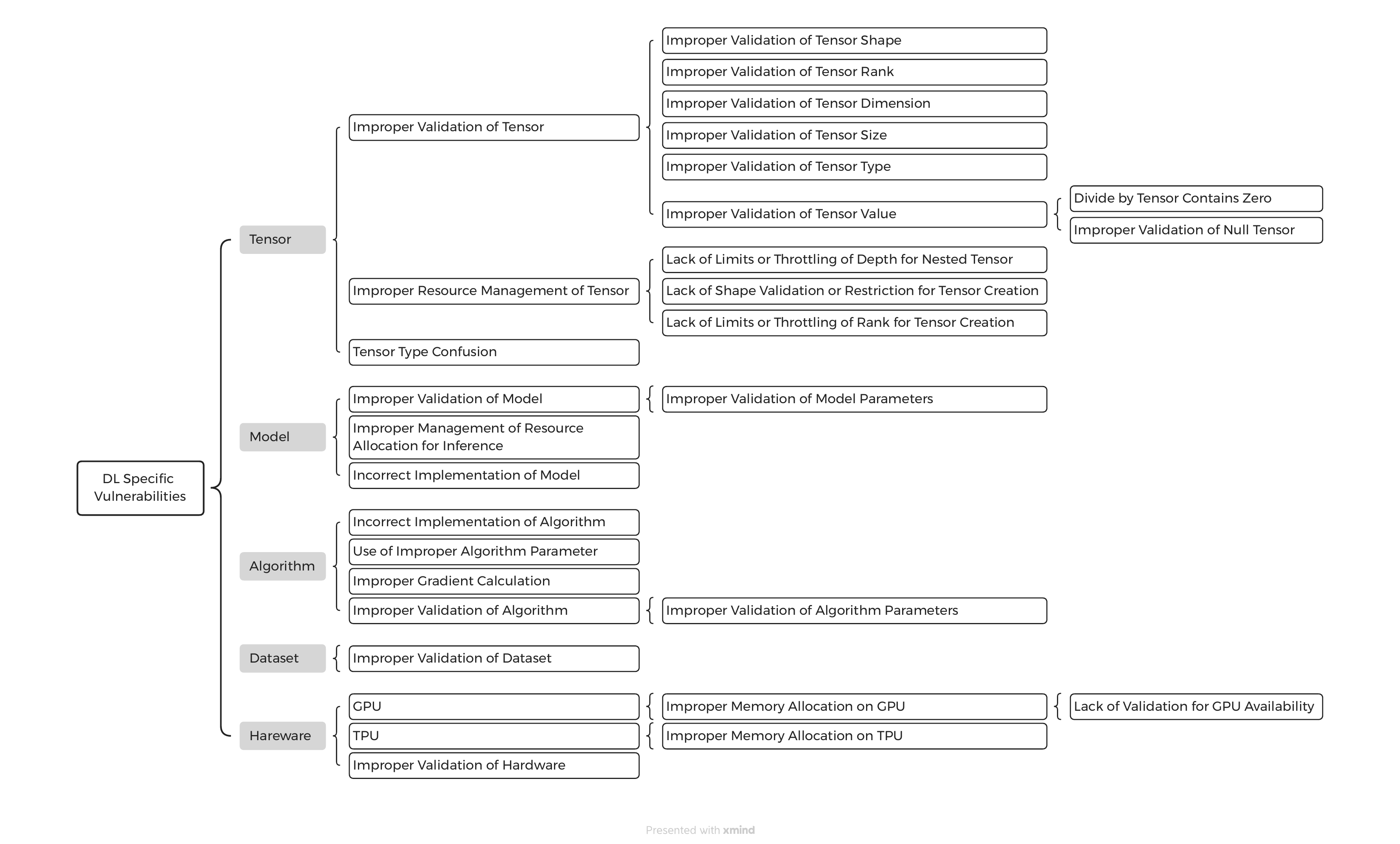}
\caption{\hl{DL specific taxonomy (CWE List additional part).}}
\label{fig:dl_specific_vulnerabilities}
\end{figure*}

\noindent \textbf{Supplementary solution for CWE List framework.} Following the challenges and findings, we observed that the current CWE List framework has certain limitations that could not guarantee the \hl{DL} systems and frameworks. 
According to Figure~\ref{fig:pillar_to_line_changed}, it is hard to distinguish dedicated efforts for the specific CWE pillar types in terms of line changes. Thus, we have also included a supplementary list of vulnerability types in this work. \hl{In addiction to the taxonomy described in Table}~\ref{tab_cwe_vuln}\hl{, we introduce the taxonomy of DL specific vulnerabilities as shown in Figure}~\ref{fig:dl_specific_vulnerabilities}.

The complete taxonomy correlations with the changed lines of code (LOC) are demonstrated in Figure~\ref{fig:taxonomy_to_line_changed}. We can see that the distribution of our dataset range evenly in terms of LOC, which are categori\hl{z}ed into `$<$10 lines', `10-99 lines', `100-999 lines', and `1000+ lines'. All vulnerability types except the top 20 are grouped into \textit{Other}. According to the result, the top 6 vulnerabilities are \textit{Improper Validation of Algorithm Parameters} (9.65\%), \textit{Improper Validation of Tensor Rank} (8.34\%), \textit{Improper Validation of Tensor Size} (5.90\%), \textit{Missing Release of Memory after Effective Lifetime} (5.48\%), \textit{Improper Restriction of Operations within the Bounds of a Memory Buffer} (5.10\%), \textit{Integer Overflow or Wraparound} (4.55\%). Three of them are the DL-related types that com\hl{e} from our proposed taxonomy. It turns out that our fine-grained DL-specific taxonomy effectively highlights the majority DL-related vulnerabilities from the conventional DL project. Accompanied by more specific definitions of DL-related vulnerability, \hl{DL} developers may easily locali\hl{z}e the DL-related vulnerability and come up with an effective patching solution. More technical details of the proposed fine-grained DL-specific taxonomy for \hl{the} CWE List \hl{are} included in~\cite{package2022investigation}.
\section{Discussion}
\label{sec:discussion_suggestions}

In this work, we systematically design a two-stream analysis framework combining a set of regular expression rules to automatically identify latent vulnerability in DL systems. With the results of the manual analysis, our framework in Figure~\ref{fig:data_collection} achieves an accuracy of 82. 39\% on GitHub and \hl{a precision of 99. 03\% on the} TensorFlow security advisory. 
It outperforms the most popular methods for empirical study~\cite{bosu2014identifying,zhou2017automated}. Furthermore, we proposed a supplementary CWE List solution \hl{originating from CWE List 4.9 covering seven pillars and a total }\typesOfVuln\hl{ different vulnerabilities with an additional taxonomy as shown in Figure}~\ref{fig:dl_specific_vulnerabilities}.

\subsection{Insights of Our Study}
\begin{figure}[t]
  \centering
  \includegraphics[width=\linewidth]{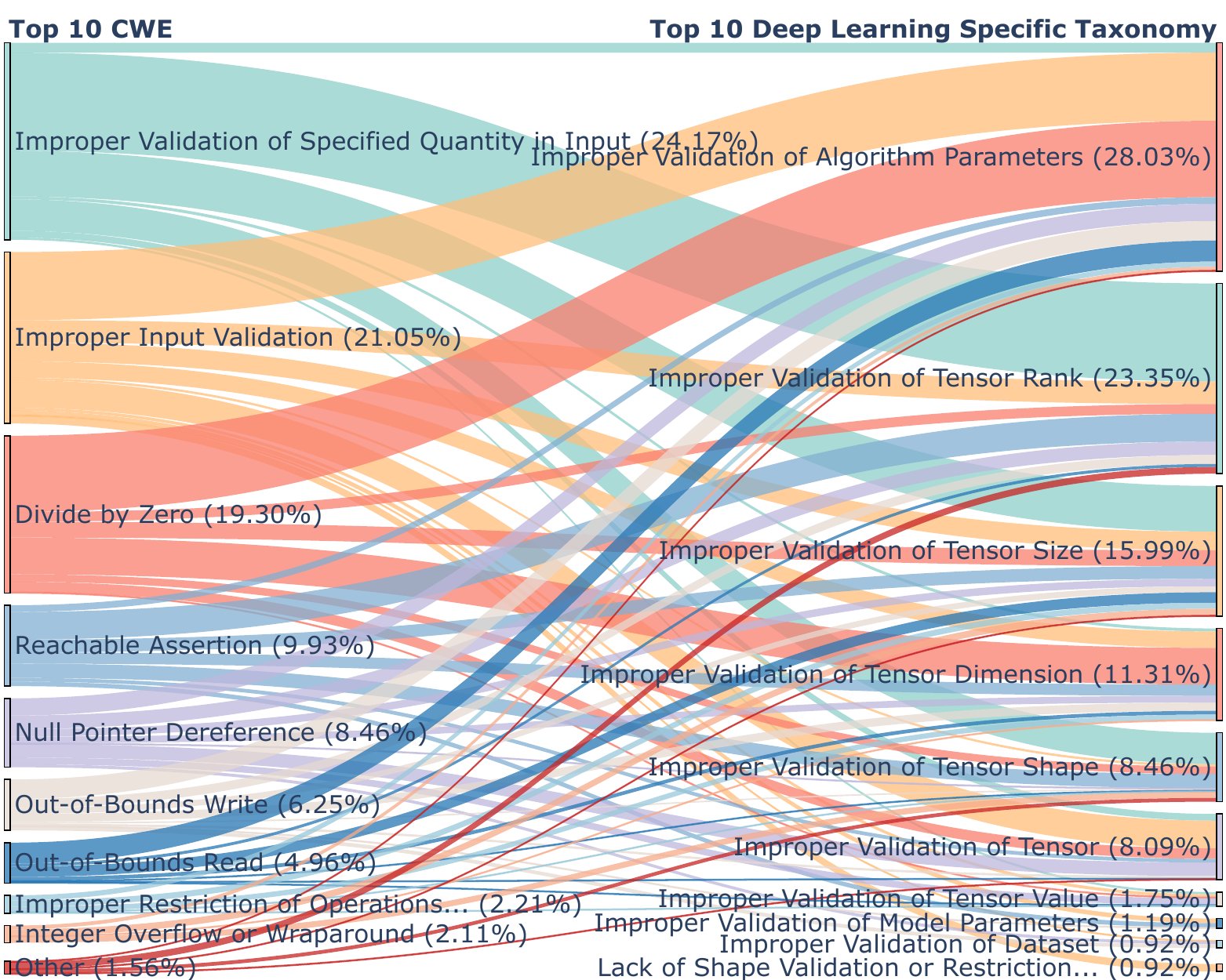}
  \caption{Top 10 CWE to top 10 DL specific taxonomy mapping.}
  \label{fig:cwe_to_dl_taxonomy}

\end{figure}

\noindent \textbf{DL-specific vulnerabilities classification.~}
In addition to the CWE 4.9 classification, we further classify the DL-specific vulnerabilities into DL-specific concepts such as \textit{Tensor}, \textit{Model}, \textit{Algorithm}, \textit{Dataset}, and \textit{Hardware (GPU,TPU)} as shown in Figure~\ref{fig:dl_specific_vulnerabilities}. Figure~\ref{fig:cwe_to_dl_taxonomy} demonstrates the mapping of the CWE List classification to \hl{the} corresponding DL-related taxonomy. It provides further insight into vulnerability in the \hl{DL} context:

\begin{itemize}[leftmargin=*]
\item \textit{Improper Validation of Algorithm Parameters} (28.03\%) and \textit{Improper Validation of Tensor (Size, Dimension, Rank, Shape, Value, Type)} (68.95\%) are the main causes covering over 96.98\% DL-related vulnerabilities in total.
\item Major DL-related vulnerabilities from current CWE List are \textit{Improper Validation of Specified Quantity in Input} (24.17\%), \textit{Improper Input Validation} (21.05\%), \textit{Divide by Zero} (19.30\%),  \textit{Reachable Assertion} (9.93\%), \textit{Null Pointer Dereference} (8.46\%), \textit{Out-of-Bounds Write} (6.25\%), \textit{Out-of-Bounds Read} (4.96\%), covering 94.12\% vulnerabilities.
\end{itemize}
Dissecting the vulnerabilities in a DL context offers a deeper understanding of the root cause and \hl{the} corresponding fix pattern for these issues. As shown in Figure~\ref{fig:cwe_to_dl_taxonomy}, improper validation of algorithm parameters (\eg batch size, channel size, filter size, input or output shape) and tensor properties (\eg size, dimension, rank, shape, value and type) are the leading causes of DL related vulnerability. Accordingly, validating these parameters and tensor properties during development can neutralize the potential issues to the maximum extent (\hl{more than} 96. 98\% of DL-related problems).

Another finding from our empirical study is that many patches contain only information on vulnerabilities symptom without pointing out the cause and \hl{the} fix method. However, the lack of description for the cause and fix method will make the patch challenging to \hl{review and evaluate} by authors or reviewers from communities. This could potentially reduce patch quality and even \hl{cause} new problems. A suggestion for improving coding quality is to clarify the symptom of the vulnerability, the cause of the issue, and how the patch fixes the problem.

\begin{figure}[t]
    \centering
    \includegraphics[width=\linewidth]{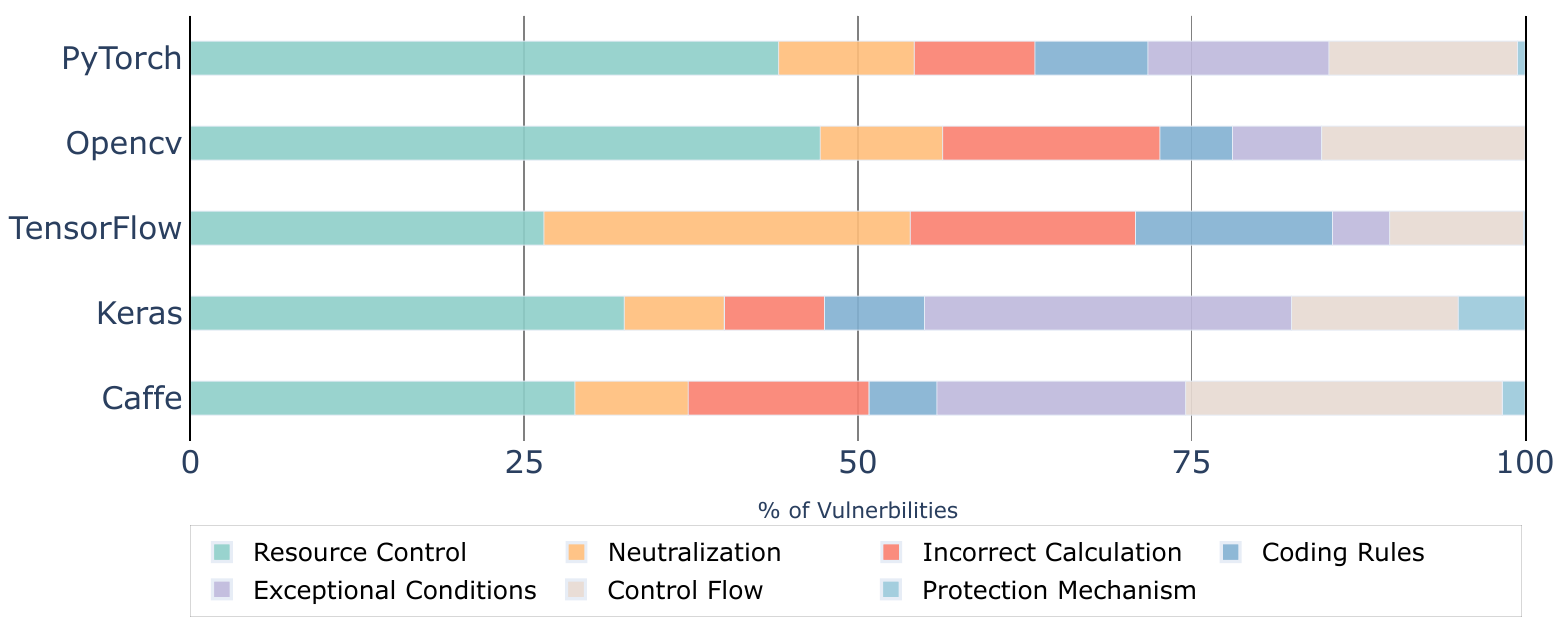}
    \caption{Percentage of vulnerabilities in each framework.}
    \label{fig:vuln_pct}
\end{figure}

\begin{figure}[t]
    \centering
    \includegraphics[width=\linewidth]{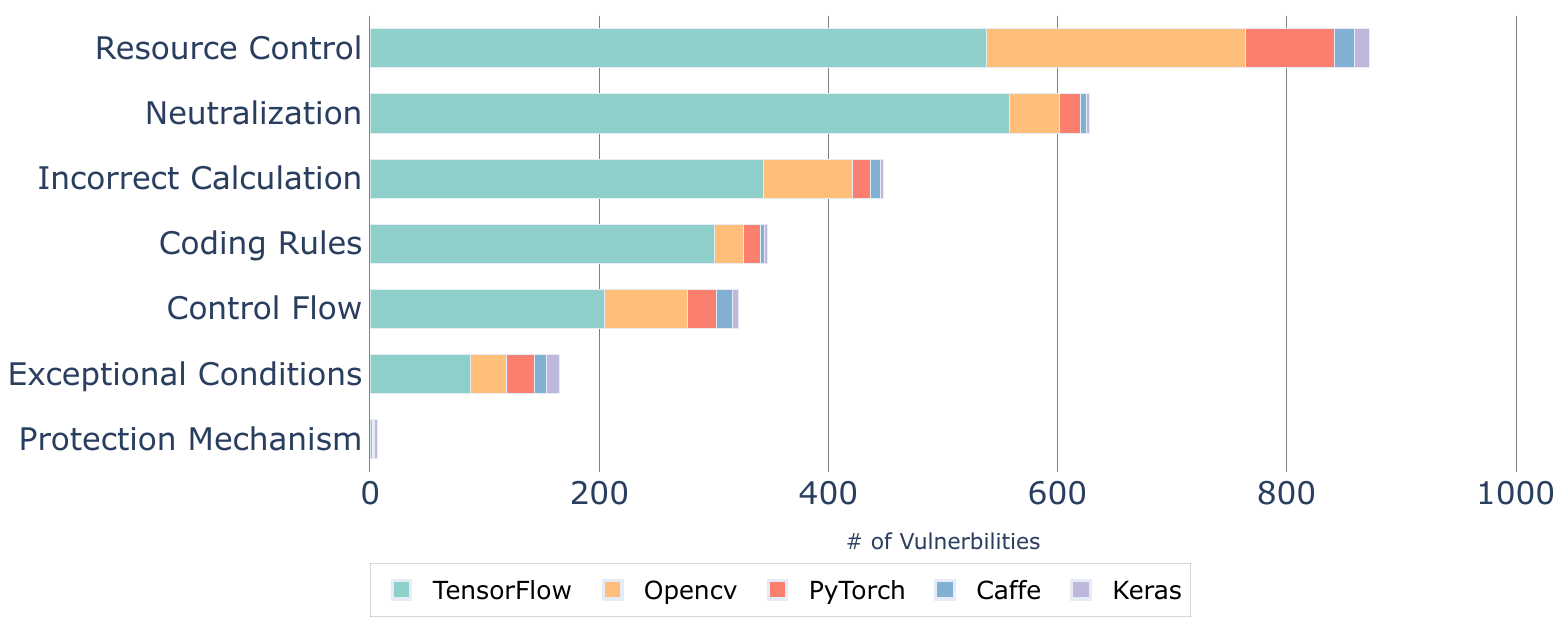}
    \caption{Distribution of vulnerabilities in each CWE pillar.}
    \label{fig:pillar_count}
\end{figure}

\noindent \hl{\textbf{Weaknesses comparison across DL frameworks.~}}
\hl{The cross-checking of vulnerabilities among five projects is illustrated in Figures}~\ref{fig:vuln_pct}\hl{ and}~\ref{fig:pillar_count}.\hl{ These vulnerabilities are categorized into seven pillars according to the CWE List. The pillars include: CWE-664 \textit{Resource Control}, CWE-682 \textit{Incorrect Calculation}, CWE-691 \textit{Control Flow}, CWE-693 \textit{Protection Mechanism}, CWE-703 \textit{Exceptional Conditions}, CWE-707 \textit{Neutralization}, and CWE-710 \textit{Coding Rules}. Based on the results of the comparison, the} resource control issue is the most common weakness in DL frameworks. TensorFlow and OpenCV handle exceptional conditions that are vastly superior to others, while Keras faces a large portion of exceptional issues. Also, TensorFlow and OpenCV almost have no protection mechanism issue. All frameworks but TensorFlow only have a small portion of the calculation vulnerability. These phenomena indicate that DL frameworks are more vulnerable to resource-related weakness\hl{es}. Moreover, TensorFlow and OpenCV might have a better code design and mechanisms to handle exceptions. Due to \hl{the fact that} TensorFlow involves massive tensor operations and algorithm executions, it is more vulnerable to numerical weaknesses.\par

Unlike TensorFlow and OpenCV, other frameworks of Keras, Caffe, and PyTorch have little attention to vulnerability issues. Although in our dataset, no CVE record is returned for these frameworks, our vulnerability search results show considerable weaknesses in the repositories. A vulnerability might not directly influence the system's usability, but malicious users can mislead \hl{the} system for irreversible consequences. The growing applications of DL systems in production indicate a new futuristic software form. Thus, we echo the raised concerns on better understanding and managing the DL systems vulnerability~\cite{spring2020managing,harzevili2022characterizing}. \hl{In particular,} documentation of well-maintained securities advisories such as TensorFlow could be a practical move. Emphasi\hl{z}ing and improving \hl{the} securities requirements for all community contributors is also suggested.\par

\noindent \textbf{Actionable findings.~}
According to the study, following actionable implications are concluded to assist developers and researchers to avoid, handle and mitigate DL vulnerabilities: 
\vspace{-1mm}
\begin{itemize}[leftmargin=*]
\item To avoid vulnerability:
\begin{itemize}[leftmargin=*]
\item Developers should properly validate the DL-incorporated resources and the defined elements in code, including the GPU availability, algorithm parameters, tensor value, type, shape, rank, and dimension. \hl{(Root cause}~\hyperref[subsubsection:root_cause_1]{1}\hl{)}
\item Developers should properly define the operation to the resources, in which of the DL systems the major concerns are from data and memory buffer. \hl{(Root cause}~\hyperref[subsubsection:root_cause_1]{1}\hl{)}
\item Developers should properly handle the range of resources in DL (Double Free, Use After Free, Out-of-Bounds). \hl{(Root cause}~\hyperref[subsubsection:root_cause_1]{1}\hl{)}
\item Developers should pay attention to the calculation (wrong calculation, integer underflow/overflow) as it is sensitive to all the DL training and testing stages. \hl{(Root cause}~\hyperref[subsubsection:root_cause_2]{2}\hl{)}
\item Developers should consider the race condition carefully, as DL projects often involve parallel computing processes using shared GPU, threads, and memory. \hl{(Root cause}~\hyperref[subsubsection:root_cause_3]{3}\hl{)}
\item Developers should provide sufficient defense against a protection mechanism for adversary attacks. \hl{(Root cause}~\hyperref[subsubsection:root_cause_4]{4}\hl{)}
\item Developers should pay more attention to the corner cases for DL systems. \hl{(Root cause}~\hyperref[subsubsection:root_cause_5]{5}\hl{)}
\item Developers should validate the type and quantity of input for neutralization issues. \hl{(Root cause}e~\hyperref[subsubsection:root_cause_6]{6}\hl{)}
\item Future work may target generalizing DL development coding standards and rules. \hl{(Root cause}~\hyperref[subsubsection:root_cause_7]{7}\hl{)}
\end{itemize}
\item For dealing with DL vulnerability detection:
\begin{itemize}[leftmargin=*]
\item Future research may target a general test oracle of DL systems testing and achieve automation to deliver the testing process. DL experts can play a pivotal role in testing design and provide essential definition criteria. \hl{(Challenge}~\hyperref[subsubsection:challenge_1]{1}\hl{)}
\item Developers should have sufficient knowledge about DL-related hardware to detect hardware-related problems \hl{(Challenge}~\hyperref[subsubsection:challenge_2]{2}\hl{)}
\item Developers should have sufficient knowledge of algorithms to improve the implementation quality and reduce the chance of undefined behavior. \hl{(Challenge}~\hyperref[subsubsection:challenge_3]{3}\hl{)}
\item Developers should understand the third-party libraries' code more deeply. \hl{(Challenge}~\hyperref[subsubsection:challenge_4]{4}\hl{)}
\end{itemize}
\item For patching the vulnerabilities:
\begin{itemize}[leftmargin=*]
\item Future research could target to improve CWE List in terms of DL-context to better localize the vulnerability. \hl{(Challenge}~\hyperref[subsubsection:challenge_5]{5}\hl{)}
\item Adding unit tests for DL is nontrivial. Finding an efficient way to conduct unit tests for DL systems could be a potential research point. \hl{(Challenge}~\hyperref[subsubsection:challenge_6]{6}\hl{)}
\item Developers can document the cause, patch methods, and symptoms in detail for reviewers to reproduce the vulnerability and verify the patching quality. \hl{(Challenge}~\hyperref[subsubsection:challenge_6]{6}\hl{)}
\item Open-source the project can bring in community support and help to detect and verify the issues. \hl{(Challenge}~\hyperref[subsubsection:challenge_6]{6}\hl{)}
\end{itemize}
\end{itemize}

\hl{Compared to general software systems, avoiding, detecting and patching vulnerabilities in DL systems presents a unique challenge due to the complex nature of DL architectures and the intricate interaction between software and hardware resources. For example, DL systems are highly sensitive to resource management issues, such as out-of-bound errors and race conditions, due to parallel processing and shared resources. Furthermore, the complexity of DL models means that even minor calculation errors can significantly impact the outcomes. In addition, as DL systems, they require the ability to defend against DL-specific vulnerabilities, such as adversarial attacks. Overcoming these challenges requires DL system developers and researchers to possess a comprehensive set of skills and knowledge, including a deep understanding of DL principles, familiarity with hardware, and proficiency in handling high-dimensional tensors and complex DL algorithms. In addition, the rapid development nature of DL technology requires ongoing learning and adaptability, raising the requirement of skills and knowledge to deal with vulnerabilities in DL systems.}

\noindent \hl{\textbf{Method versatility.} As shown in Figure}~\ref{fig:data_collection}\hl{, Our innovative two-stream analysis framework leverages data from the NVD and GitHub (PRs, Issues, and Commits), ensuring easy access and adaptability to various open-source projects. Furthermore, our analysis framework integrates DL-specific concepts, such as tensors, models, algorithms, datasets, and hardware. This integration provides a more nuanced, clear, and precise classification of DL-specific vulnerabilities. Combining a generalized analysis framework with a DL-specific taxonomy gives our analysis framework superior versatility and seamless extensibility to a wide variety of DL systems, including transformer and large-scale language models (LLMs).}

\subsection{Threats to Validity}
\label{sec:threats_to_validity}
\noindent \textbf{DL Framework Selection.~}
In our study, we have included the five most popular \hl{DL} projects. While the main goal is to represent the vulnerabilities in \hl{DL} systems comprehensively, our dataset may not cover all existing DL frameworks, in particular, we observe that TensorFlow represents the most vulnerabilities at this stage. In addition, \hl{DL} frameworks such as CNTK and deeplearning4j are not included due to limited source code or data availability. \hl{In particular,} we carefully select the top five \hl{DL} projects according to GitHub stars to mitigate threats, which provides us \hl{with} a raw dataset of 73,215 closed issues. To facilitate the investigation of vulnerabilities in DL systems, we have released the replication package of our work. We anticipate that future investigation will be covered with more projects.

\noindent \textbf{Vulnerability identification.~}
To overcome the challenge of the availability of vulnerabilities, we followed previous studies to collect PR, issues, and commits. However, the vulnerability identification process may not be sufficient and can miss some real vulnerability of significant impacts in DL systems. Although the research methodology is similar in vulnerability collection for \textit{machine learning models for vulnerability detection} task, we have conducted considerable efforts on identifying the vulnerabilities with an audit process. Future work will be extended with refined regular expressions for searching.

\noindent \textbf{Manual labeling process.~}
While two authors are actively involved for manual analysis, the results could be prone to subjective opinions. In order to mitigate this threat, statistic\hl{al} analysis and substantial discussions are conducted for the classification and labeling process. \hl{Although} the work is managed independently, the authors support each other with good findings to build the taxonomy. All of the results with the working code, dataset and scripts are released for further investigation~\cite{package2022investigation}.
\section{Conclusion}
\label{sec:conclusion}
In this work, we present a first \textit{systematic} study \hl{of its kind to better understand the vulnerabilities of software} in DL systems. \hl{In particular}, we have \hl{innovatively} designed a two-stream \hl{analysis} framework to \hl{address the data source concern, which} provide\hl{s} the findings of such root causes and challenges in the detection and patch process. Ahead of the thought experiments for vulnerability management in artificial intelligence systems, our empirical study also includes a supplementary CWE List proposition to provide much more \hl{insights}. Following the findings, a summary of actionable implications is provided for security practice in DL related projects. While we do not provide strong recommendations for or against, in which the changes for vulnerability study can be nuanced and also complicated, we believe \hl{that} this work has bridged the gap and provided a solid basis for future work to build a secure guarantee of DL systems.
\bibliography{IEEEabrv,main}

\begin{thebibliography}{10}
\providecommand{\url}[1]{#1}
\csname url@samestyle\endcsname
\providecommand{\newblock}{\relax}
\providecommand{\bibinfo}[2]{#2}
\providecommand{\BIBentrySTDinterwordspacing}{\spaceskip=0pt\relax}
\providecommand{\BIBentryALTinterwordstretchfactor}{4}
\providecommand{\BIBentryALTinterwordspacing}{\spaceskip=\fontdimen2\font plus
\BIBentryALTinterwordstretchfactor\fontdimen3\font minus \fontdimen4\font\relax}
\providecommand{\BIBforeignlanguage}[2]{{%
\expandafter\ifx\csname l@#1\endcsname\relax
\typeout{** WARNING: IEEEtran.bst: No hyphenation pattern has been}%
\typeout{** loaded for the language `#1'. Using the pattern for}%
\typeout{** the default language instead.}%
\else
\language=\csname l@#1\endcsname
\fi
#2}}
\providecommand{\BIBdecl}{\relax}
\BIBdecl

\bibitem{liu2020using}
J.~Liu, Q.~Huang, X.~Xia, E.~Shihab, D.~Lo, and S.~Li, ``Is using deep learning frameworks free? characterizing technical debt in deep learning frameworks,'' in \emph{Proceedings of the ACM/IEEE 42nd International Conference on Software Engineering: Software Engineering in Society}, 2020, pp. 1--10.

\bibitem{chen2021empirical}
Z.~Chen, H.~Yao, Y.~Lou, Y.~Cao, Y.~Liu, H.~Wang, and X.~Liu, ``An empirical study on deployment faults of deep learning based mobile applications,'' in \emph{2021 IEEE/ACM 43rd International Conference on Software Engineering (ICSE)}.\hskip 1em plus 0.5em minus 0.4em\relax IEEE, 2021, pp. 674--685.

\bibitem{chen2020comprehensive}
Z.~Chen, Y.~Cao, Y.~Liu, H.~Wang, T.~Xie, and X.~Liu, ``A comprehensive study on challenges in deploying deep learning based software,'' in \emph{Proceedings of the 28th ACM Joint Meeting on European Software Engineering Conference and Symposium on the Foundations of Software Engineering}, 2020, pp. 750--762.

\bibitem{lou2020understanding}
Y.~Lou, Z.~Chen, Y.~Cao, D.~Hao, and L.~Zhang, ``Understanding build issue resolution in practice: symptoms and fix patterns,'' in \emph{Proceedings of the 28th ACM Joint Meeting on European Software Engineering Conference and Symposium on the Foundations of Software Engineering}, 2020, pp. 617--628.

\bibitem{wu2021empirical}
D.~Wu, B.~Shen, and Y.~Chen, ``An empirical study on tensor shape faults in deep learning systems,'' \emph{arXiv preprint arXiv:2106.02887}, 2021.

\bibitem{knight2002safety}
J.~C. Knight, ``Safety critical systems: challenges and directions,'' in \emph{Proceedings of the 24th international conference on software engineering}, 2002, pp. 547--550.

\bibitem{wan2021machine}
C.~Wan, S.~Liu, H.~Hoffmann, M.~Maire, and S.~Lu, ``Are machine learning cloud apis used correctly?'' in \emph{2021 IEEE/ACM 43rd International Conference on Software Engineering (ICSE)}.\hskip 1em plus 0.5em minus 0.4em\relax IEEE, 2021, pp. 125--137.

\bibitem{zhang2021unveiling}
Z.~Zhang, Y.~Yang, X.~Xia, D.~Lo, X.~Ren, and J.~Grundy, ``Unveiling the mystery of api evolution in deep learning frameworks a case study of tensorflow 2,'' in \emph{2021 IEEE/ACM 43rd International Conference on Software Engineering: Software Engineering in Practice (ICSE-SEIP)}.\hskip 1em plus 0.5em minus 0.4em\relax IEEE, 2021, pp. 238--247.

\bibitem{islam2019comprehensive}
M.~J. Islam, G.~Nguyen, R.~Pan, and H.~Rajan, ``A comprehensive study on deep learning bug characteristics,'' in \emph{Proceedings of the 2019 27th ACM Joint Meeting on European Software Engineering Conference and Symposium on the Foundations of Software Engineering}, 2019, pp. 510--520.

\bibitem{jia2020tfbugs}
L.~Jia, H.~Zhong, X.~Wang, L.~Huang, and X.~Lu, ``An empirical study on bugs inside tensorflows,'' in \emph{Proc. DASFAA}, 2020, pp. 604--620.

\bibitem{yan2021exposing}
M.~Yan, J.~Chen, X.~Zhang, L.~Tan, G.~Wang, and Z.~Wang, ``Exposing numerical bugs in deep learning via gradient back-propagation,'' in \emph{Proceedings of the 29th ACM Joint Meeting on European Software Engineering Conference and Symposium on the Foundations of Software Engineering}, 2021, pp. 627--638.

\bibitem{kim2021denchmark}
M.~Kim, Y.~Kim, and E.~Lee, ``Denchmark: A bug benchmark of deep learning-related software,'' in \emph{2021 IEEE/ACM 18th International Conference on Mining Software Repositories (MSR)}.\hskip 1em plus 0.5em minus 0.4em\relax IEEE, 2021, pp. 540--544.

\bibitem{cao2021characterizing}
J.~Cao, B.~Chen, C.~Sun, L.~Hu, and X.~Peng, ``Characterizing performance bugs in deep learning systems,'' \emph{arXiv preprint arXiv:2112.01771}, 2021.

\bibitem{nejadgholi2019study}
M.~Nejadgholi and J.~Yang, ``A study of oracle approximations in testing deep learning libraries,'' in \emph{2019 34th IEEE/ACM International Conference on Automated Software Engineering (ASE)}.\hskip 1em plus 0.5em minus 0.4em\relax IEEE, 2019, pp. 785--796.

\bibitem{pham2019cradle}
H.~V. Pham, T.~Lutellier, W.~Qi, and L.~Tan, ``Cradle: cross-backend validation to detect and localize bugs in deep learning libraries,'' in \emph{2019 IEEE/ACM 41st International Conference on Software Engineering (ICSE)}.\hskip 1em plus 0.5em minus 0.4em\relax IEEE, 2019, pp. 1027--1038.

\bibitem{wang2020deep}
Z.~Wang, M.~Yan, J.~Chen, S.~Liu, and D.~Zhang, ``Deep learning library testing via effective model generation,'' in \emph{Proceedings of the 28th ACM Joint Meeting on European Software Engineering Conference and Symposium on the Foundations of Software Engineering}, 2020, pp. 788--799.

\bibitem{wu2019towards}
D.~Wu, D.~Gao, E.~K. Cheng, Y.~Cao, J.~Jiang, and R.~H. Deng, ``Towards understanding android system vulnerabilities: Techniques and insights,'' in \emph{Proceedings of the 2019 ACM Asia Conference on Computer and Communications Security}, 2019, pp. 295--306.

\bibitem{abadi2016tensorflow}
M.~Abadi, P.~Barham, J.~Chen, Z.~Chen, A.~Davis, J.~Dean, M.~Devin, S.~Ghemawat, G.~Irving, M.~Isard \emph{et~al.}, ``$\{$TensorFlow$\}$: A system for $\{$Large-Scale$\}$ machine learning,'' in \emph{12th USENIX symposium on operating systems design and implementation (OSDI 16)}, 2016, pp. 265--283.

\bibitem{jia2014caffe}
Y.~Jia, E.~Shelhamer, J.~Donahue, S.~Karayev, J.~Long, R.~Girshick, S.~Guadarrama, and T.~Darrell, ``Caffe: Convolutional architecture for fast feature embedding,'' in \emph{Proceedings of the 22nd ACM international conference on Multimedia}, 2014, pp. 675--678.

\bibitem{bradski2008learning}
G.~Bradski and A.~Kaehler, \emph{Learning OpenCV: Computer vision with the OpenCV library}.\hskip 1em plus 0.5em minus 0.4em\relax " O'Reilly Media, Inc.", 2008.

\bibitem{paszke2019pytorch}
A.~Paszke, S.~Gross, F.~Massa, A.~Lerer, J.~Bradbury, G.~Chanan, T.~Killeen, Z.~Lin, N.~Gimelshein, L.~Antiga \emph{et~al.}, ``Pytorch: An imperative style, high-performance deep learning library,'' \emph{Advances in neural information processing systems}, vol.~32, 2019.

\bibitem{gulli2017deep}
A.~Gulli and S.~Pal, \emph{Deep learning with Keras}.\hskip 1em plus 0.5em minus 0.4em\relax Packt Publishing Ltd, 2017.

\bibitem{package2022investigation}
Z.~Lai, H.~Chen, R.~Sun, Y.~Zhang, M.~Xue, and D.~Yuan, ``Replication package,'' \url{https://github.com/codelzz/Vulnerabilities4DLSystem}.

\bibitem{rao2018deep}
Q.~Rao and J.~Frtunikj, ``Deep learning for self-driving cars: Chances and challenges,'' in \emph{Proceedings of the 1st International Workshop on Software Engineering for AI in Autonomous Systems}, 2018, pp. 35--38.

\bibitem{liu2020fned}
Y.~Liu and Y.-F.~B. Wu, ``Fned: a deep network for fake news early detection on social media,'' \emph{ACM Transactions on Information Systems (TOIS)}, vol.~38, no.~3, pp. 1--33, 2020.

\bibitem{wang2020deep1}
J.~Wang, K.~Sun, T.~Cheng, B.~Jiang, C.~Deng, Y.~Zhao, D.~Liu, Y.~Mu, M.~Tan, X.~Wang \emph{et~al.}, ``Deep high-resolution representation learning for visual recognition,'' \emph{IEEE transactions on pattern analysis and machine intelligence}, vol.~43, no.~10, pp. 3349--3364, 2020.

\bibitem{cysneiros2018software}
L.~M. Cysneiros, M.~Raffi, and J.~C.~S. do~Prado~Leite, ``Software transparency as a key requirement for self-driving cars,'' in \emph{2018 IEEE 26th international requirements engineering conference (RE)}.\hskip 1em plus 0.5em minus 0.4em\relax IEEE, 2018, pp. 382--387.

\bibitem{bosch2021engineering}
J.~Bosch, H.~H. Olsson, and I.~Crnkovic, ``Engineering ai systems: A research agenda,'' in \emph{Artificial Intelligence Paradigms for Smart Cyber-Physical Systems}.\hskip 1em plus 0.5em minus 0.4em\relax IGI global, 2021, pp. 1--19.

\bibitem{nikanjam2021design}
A.~Nikanjam and F.~Khomh, ``Design smells in deep learning programs: An empirical study,'' in \emph{2021 IEEE International Conference on Software Maintenance and Evolution (ICSME)}.\hskip 1em plus 0.5em minus 0.4em\relax IEEE, 2021, pp. 332--342.

\bibitem{giray2021software}
G.~Giray, ``A software engineering perspective on engineering machine learning systems: State of the art and challenges,'' \emph{Journal of Systems and Software}, vol. 180, p. 111031, 2021.

\bibitem{de2019understanding}
E.~de~Souza~Nascimento, I.~Ahmed, E.~Oliveira, M.~P. Palheta, I.~Steinmacher, and T.~Conte, ``Understanding development process of machine learning systems: Challenges and solutions,'' in \emph{2019 ACM/IEEE International Symposium on Empirical Software Engineering and Measurement (ESEM)}.\hskip 1em plus 0.5em minus 0.4em\relax IEEE, 2019, pp. 1--6.

\bibitem{john2020architecting}
M.~M. John, H.~Holmstr{\"o}m~Olsson, and J.~Bosch, ``Architecting ai deployment: A systematic review of state-of-the-art and state-of-practice literature,'' in \emph{International Conference on Software Business}.\hskip 1em plus 0.5em minus 0.4em\relax Springer, 2020, pp. 14--29.

\bibitem{ye2020put}
W.~Ye, R.~Hu, and M.~Enev, ``Put deep learning to work: Accelerate deep learning through amazon sagemaker and ml services,'' in \emph{Proceedings of the 26th ACM SIGKDD international conference on knowledge discovery \& data mining}, 2020, pp. 3496--3496.

\bibitem{malta2019exploring}
E.~M. Malta, S.~Avila, and E.~Borin, ``Exploring the cost-benefit of aws ec2 gpu instances for deep learning applications,'' in \emph{Proceedings of the 12th IEEE/ACM International Conference on Utility and Cloud Computing}, 2019, pp. 21--29.

\bibitem{papernot2018marauder}
N.~Papernot, ``A marauder's map of security and privacy in machine learning: An overview of current and future research directions for making machine learning secure and private,'' in \emph{Proceedings of the 11th ACM Workshop on Artificial Intelligence and Security}, 2018, pp. 1--1.

\bibitem{he2020towards}
Y.~He, G.~Meng, K.~Chen, X.~Hu, and J.~He, ``Towards security threats of deep learning systems: A survey,'' \emph{IEEE Transactions on Software Engineering}, 2020.

\bibitem{evtimov2020security}
I.~Evtimov, W.~Cui, E.~Kamar, E.~Kiciman, T.~Kohno, and J.~Li, ``Security and machine learning in the real world,'' \emph{arXiv preprint arXiv:2007.07205}, 2020.

\bibitem{chen2022security}
H.~Chen and M.~A. Babar, ``Security for machine learning-based software systems: a survey of threats, practices and challenges,'' \emph{arXiv preprint arXiv:2201.04736}, 2022.

\bibitem{liu2021machine}
B.~Liu, M.~Ding, S.~Shaham, W.~Rahayu, F.~Farokhi, and Z.~Lin, ``When machine learning meets privacy: A survey and outlook,'' \emph{ACM Computing Surveys (CSUR)}, vol.~54, no.~2, pp. 1--36, 2021.

\bibitem{harzevili2022characterizing}
N.~S. Harzevili, J.~Shin, J.~Wang, and S.~Wang, ``Characterizing and understanding software security vulnerabilities in machine learning libraries,'' \emph{arXiv preprint arXiv:2203.06502}, 2022.

\bibitem{cert2020}
SEI/CERT, ``Cert/cc vulnerability note vu\#425163 - machine learning classifiers trained via gradient descent are vulnerable to arbitrary misclassification attack,'' https://kb.cert.org/vuls/id/425163, Mar 2020, [Online; accessed 20-Feb-2022].

\bibitem{kim2020impact}
A.~A.~H. Kim, ``The impact of platform vulnerabilities in ai systems,'' Ph.D. dissertation, Massachusetts Institute of Technology, 2020.

\bibitem{filus2023software}
K.~Filus and J.~Doma{\'n}ska, ``Software vulnerabilities in tensorflow-based deep learning applications,'' \emph{Computers \& Security}, vol. 124, p. 102948, 2023.

\bibitem{chen2021security}
H.~Chen, Y.~Zhang, Y.~Cao, and J.~Xie, ``Security issues and defensive approaches in deep learning frameworks,'' \emph{Tsinghua Science and Technology}, vol.~26, no.~6, pp. 894--905, 2021.

\bibitem{thung2012empirical}
F.~Thung, S.~Wang, D.~Lo, and L.~Jiang, ``An empirical study of bugs in machine learning systems,'' in \emph{2012 IEEE 23rd International Symposium on Software Reliability Engineering}.\hskip 1em plus 0.5em minus 0.4em\relax IEEE, 2012, pp. 271--280.

\bibitem{zhang2020machine}
J.~M. Zhang, M.~Harman, L.~Ma, and Y.~Liu, ``Machine learning testing: Survey, landscapes and horizons,'' \emph{IEEE Transactions on Software Engineering}, vol.~48, no.~1, pp. 1--36, 2020.

\bibitem{zhang2018empirical}
Y.~Zhang, Y.~Chen, S.-C. Cheung, Y.~Xiong, and L.~Zhang, ``An empirical study on tensorflow program bugs,'' in \emph{Proceedings of the 27th ACM SIGSOFT International Symposium on Software Testing and Analysis}, 2018, pp. 129--140.

\bibitem{du2021empirical}
X.~Du, Z.~Zheng, L.~Ma, and J.~Zhao, ``An empirical study on common bugs in deep learning compilers,'' in \emph{2021 IEEE 32nd International Symposium on Software Reliability Engineering (ISSRE)}.\hskip 1em plus 0.5em minus 0.4em\relax IEEE, 2021, pp. 184--195.

\bibitem{xie2011testing}
X.~Xie, J.~W. Ho, C.~Murphy, G.~Kaiser, B.~Xu, and T.~Y. Chen, ``Testing and validating machine learning classifiers by metamorphic testing,'' \emph{Journal of Systems and Software}, vol.~84, no.~4, pp. 544--558, 2011.

\bibitem{christou2023ivysyn}
N.~Christou, D.~Jin, V.~Atlidakis, B.~Ray, and V.~P. Kemerlis, ``$\{$IvySyn$\}$: Automated vulnerability discovery in deep learning frameworks,'' in \emph{32nd USENIX Security Symposium (USENIX Security 23)}, 2023, pp. 2383--2400.

\bibitem{wang2021prioritizing}
Z.~Wang, H.~You, J.~Chen, Y.~Zhang, X.~Dong, and W.~Zhang, ``Prioritizing test inputs for deep neural networks via mutation analysis,'' in \emph{2021 IEEE/ACM 43rd International Conference on Software Engineering (ICSE)}.\hskip 1em plus 0.5em minus 0.4em\relax IEEE, 2021, pp. 397--409.

\bibitem{zhang2021predoo}
X.~Zhang, N.~Sun, C.~Fang, J.~Liu, J.~Liu, D.~Chai, J.~Wang, and Z.~Chen, ``Predoo: precision testing of deep learning operators,'' in \emph{Proceedings of the 30th ACM SIGSOFT International Symposium on Software Testing and Analysis}, 2021, pp. 400--412.

\bibitem{zhang2021duo}
X.~Zhang, J.~Liu, N.~Sun, C.~Fang, J.~Liu, J.~Wang, D.~Chai, and Z.~Chen, ``Duo: Differential fuzzing for deep learning operators,'' \emph{IEEE Transactions on Reliability}, vol.~70, no.~4, pp. 1671--1685, 2021.

\bibitem{CWE2022}
MITRE, ``Cwe list version 4.9,'' \url{https://cwe.mitre.org/data/published/cwe_v4.9.pdf}, 2022.

\bibitem{CVE2022}
------, ``Cve,'' \url{https://cve.mitre.org/}, 2022.

\bibitem{NVD2022}
NVD, ``National vulnerability database,'' \url{https://nvd.nist.gov/vuln/}, 2022.

\bibitem{bosu2014identifying}
A.~Bosu, J.~C. Carver, M.~Hafiz, P.~Hilley, and D.~Janni, ``Identifying the characteristics of vulnerable code changes: An empirical study,'' in \emph{Proceedings of the 22nd ACM SIGSOFT international symposium on foundations of software engineering}, 2014, pp. 257--268.

\bibitem{zhou2017automated}
Y.~Zhou and A.~Sharma, ``Automated identification of security issues from commit messages and bug reports,'' in \emph{Proceedings of the 2017 11th joint meeting on foundations of software engineering}, 2017, pp. 914--919.

\bibitem{githubsecurity2022}
GitHub, ``Github security advisory,'' \url{https://github.com/advisories/}, 2022.

\bibitem{holton2007coding}
J.~A. Holton, ``The coding process and its challenges,'' \emph{The Sage handbook of grounded theory}, vol.~3, pp. 265--289, 2007.

\bibitem{banerjee1999beyond}
M.~Banerjee, M.~Capozzoli, L.~McSweeney, and D.~Sinha, ``Beyond kappa: A review of interrater agreement measures,'' \emph{Canadian journal of statistics}, vol.~27, no.~1, pp. 3--23, 1999.

\bibitem{tf2017pr10298}
TensorFlow, ``Philoxrandom: Fix race in gpu fill function,'' \url{https://github.com/tensorflow/tensorflow/pull/10298}, 2017.

\bibitem{keras2017pr5049}
Keras, ``Model generators: Make sure all threads finish when stop is requested,'' \url{https://github.com/keras-team/keras/pull/5049}, 2017.

\bibitem{pytorch2018pr4857}
PyTorch, ``Don't throw exceptions inside openmp parallel block,'' \url{https://github.com/pytorch/pytorch/pull/4857}, 2018.

\bibitem{keras2018pr7071}
Keras, ``training.py \_check\_loss\_and\_target\_compatibility\(\) fix crash if y is none,'' \url{https://github.com/keras-team/keras/pull/7071}, 2018.

\bibitem{caffe2014pr1048}
Caffe, ``Conv layer: fix crash by setting weight pointer,'' \url{https://github.com/BVLC/caffe/pull/1048}, 2014.

\bibitem{tf2020pr38417}
TensorFlow, ``r2.2-rc3 cherry-pick request: Fix a bug that profile xla gpu crashes oom,'' \url{https://github.com/tensorflow/tensorflow/pull/38417}, 2020.

\bibitem{pytorch2021pr58651}
PyTorch, ``Onnx fix the issue of converting empty list to sequence,'' \url{https://github.com/pytorch/pytorch/pull/58651}, 2021.

\bibitem{pytorch2018pr8246}
------, ``Hotfix: Fix test\_cuda import in test\_cuda,'' \url{https://github.com/pytorch/pytorch/pull/8246}, 2018.

\bibitem{caffe2014pr633}
Caffe, ``Support cpu only memcpy,'' \url{https://github.com/BVLC/caffe/pull/633}, 2014.

\bibitem{pytorch2018pr5376}
PyTorch, ``Dataparallel: Gpu imbalance warning,'' \url{https://github.com/pytorch/pytorch/pull/5376}, 2018.

\bibitem{tf2020pr42615}
TensorFlow, ``Fix dynamicpartitionopgpu when running on multiple gpus,'' \url{https://github.com/tensorflow/tensorflow/pull/42615}, 2020.

\bibitem{opencv2019pr5161}
OpenCV, ``fix some functions for valid processing of empty string content,'' \url{https://github.com/opencv/opencv/pull/5161}, 2019.

\bibitem{opencv2019pr13692}
------, ``Remove assert\_any\_throw tests for myriad plugin,'' \url{https://github.com/opencv/opencv/pull/13692}, 2019.

\bibitem{tf2019pr24674}
TensorFlow, ``Tftrt: Support dilated convolutions,'' \url{https://github.com/tensorflow/tensorflow/pull/24674}, 2019.

\bibitem{tf2016pr5349}
------, ``Include libcurl into the bazel build,'' \url{https://github.com/tensorflow/tensorflow/pull/5349}, 2016.

\bibitem{tf2020pr36856}
------, ``Tflite fix for the segmentation fault. when quantize conv\_2d with dilation \!\= 1,'' \url{https://github.com/tensorflow/tensorflow/pull/36856}, 2020.

\bibitem{tf2020pr36692}
------, ``Tfliteconverter\: segmentation fault when dilation is not (1,1) in conv\_2d,'' \url{https://github.com/tensorflow/tensorflow/issues/36692}, 2020.

\bibitem{pytorch2018pr5585}
PyTorch, ``Fix memory leak when using multiple workers on window,'' \url{https://github.com/pytorch/pytorch/pull/5585}, 2018.

\bibitem{spring2020managing}
J.~M. Spring, A.~Galyardt, A.~D. Householder, and N.~VanHoudnos, ``On managing vulnerabilities in ai/ml systems,'' in \emph{New Security Paradigms Workshop 2020}, 2020, pp. 111--126.

\end{thebibliography}
\bibliographystyle{IEEEtran}


\end{document}